\documentclass[reprint,nofootinbib,amsmath,amssymb,showkeys,aps]{revtex4-1}

\usepackage{amssymb}
\usepackage{array,multirow}
\usepackage{enumerate}
\usepackage{bm}
\usepackage{amsmath}
\usepackage{graphicx}

\usepackage[flushleft]{threeparttable}

\begin{document}

\title{Measuring the sound horizon and absolute magnitude of SNIa by maximizing the consistency between low-redshift data sets}

\author{Adri\`a G\'omez-Valent$^{1,2}$}\email{agvalent@roma2.infn.it}

\affiliation{$^1$ Dipartimento di Fisica, Università di Roma Tor Vergata, via della Ricerca Scientifica, 1, 00133, Roma, Italy}
\affiliation{$^2$ INFN, Sezione di Roma 2, Università di Roma Tor Vergata, via della Ricerca Scientifica, 1, 00133 Roma, Italy}

\begin{abstract}
The comoving sound horizon at the baryon drag epoch, $r_{d}$, encapsulates very important physical information about the pre-recombination era and serves as a cosmic standard ruler. On the other hand, the absolute magnitude of supernovae of Type Ia (SNIa), $M$, is pivotal to infer the distances to these standard candles. Having access to (at least) one of these two quantities is crucial to measure the Hubble parameter $H_0$ from BAO/SNIa data. In this work we present a new method to measure how long is the cosmic ruler and how bright are the standard candles independently from the main drivers of the $H_0$ tension, namely by avoiding (i) the use of CMB data; (ii) the calibration of SNIa in the first steps of the cosmic distance ladder; and (iii) the assumption of any concrete cosmological model. We only assume that SNIa can be safely employed as standard candles and $r_d$ as a standard ruler, together with the validity of the Cosmological Principle and the metric description of gravity, with photons propagating in null geodesics and the conservation of the photon number. Our method is based on the minimization of a loss function that represents the level of inconsistency between the low-redshift data sets employed in this study, to wit: SNIa, BAO and cosmic chronometers. In our main analysis we obtain: $r_d=(146.0^{+4.2}_{-5.1})$ Mpc, $M=-19.362^{+0.078}_{-0.067}$. The former is fully compatible with {\it Planck}'s $\Lambda$CDM best-fit cosmology, but still leaves plenty of room for new physics before the decoupling, whereas our constraint on $M$ lies closer to the value preferred by the concordance model, although it is only $\sim 1.4\sigma$ below the SH0ES measurement.  
\end{abstract}

\keywords{Cosmology: observations -- Cosmology: theory -- cosmological parameters -- Large-Scale Structure of Universe}

\maketitle


\section{Introduction}\label{sec:intro}

How old is the universe? How distant are the cosmological objects that we detect with our telescopes? Our answers to these questions depend crucially on how accurately we can measure the current universe's expansion rate, i.e. the Hubble parameter $H_0$, since it sets both, the time and distance scales in cosmology. Now, after more than nine decades from its first measurement \citep{Hubble:1929} several observational sources tell us that its value should fall in the range $H_0\in (65,75)$ km/s/Mpc. However, there is a $5.0\sigma$ tension between the value inferred for this parameter with the classical cosmic distance ladder by the SH0ES team, $H_0=(73.04\pm 1.01)$ km/s/Mpc \citep{Riess:2021jrx}, which is almost fully cosmology-independent, and the one inferred with the cosmic microwave background (CMB) data by the {\it Planck} Collaboration, under the assumption of the flat $\Lambda$CDM model, $H_0=(67.36\pm 0.54)$ km/s/Mpc \citep{Planck:2018vyg}. SH0ES measures in the first steps of the ladder a somewhat higher value of the absolute magnitude of supernovae of Type Ia (SNIa), $M$, than the one preferred by the CMB best-fit standard model, and this is what ultimately triggers the $H_0$ tension, see e.g. \citep{Camarena:2019moy,Efstathiou:2021ocp}. Other observational teams, though, do not find any substantial discrepancy \citep{Freedman:2019jwv,Freedman:2021ahq}, and it is still possible that systematic errors play an important role in this story \citep{Efstathiou:2020wxn,Mortsell:2021nzg}. Nevertheless, the tension has persisted and grown in a consistent way in the last years thanks to the gain in precision of the modern observational facilities employed to explore the local universe \citep{Riess:2009pu,Riess:2016jrr,Riess:2020fzl} and measure the CMB anisotropies \citep{WMAP:2012nax,Planck:2015fie}. Although there exist some internal tensions at the $\sim 2\sigma$ level in {\it Planck}'s data (e.g. between the $\Lambda$CDM best-fit parameters obtained from low and high multipoles or the amount of lensing in the TT spectrum), their results seem to be consistent with other CMB experiments, as WMAP \citep{WMAP:2012nax}, SPT \citep{SPT:2017sjt} and ACT \citep{ACT:2020gnv}. Moreover, constraints from baryon acoustic oscillations (BAO) and big bang nucleosynthesis, independent from CMB, lead to values of the Hubble parameter lying again in the lower range in the context of the standard model, more in accordance with {\it Planck} \citep{Addison:2017fdm,Cuceu:2019for}. This is also found when use is made of the inverse cosmic distance ladder \citep{Aubourg:2014yra,Cuesta:2014asa,Feeney:2018mkj,Camarena:2019rmj}, which assumes again standard pre-recombination (and most of the cases also standard late-time) physics. Interestingly, standard sirens \citep{LIGOScientific:2017adf,Palmese:2021mjm} and strongly lensed quasars \citep{Denzel:2020zuq} also allow us to measure $H_0$, but they are still not able to arbitrate the tension. 

Cosmological models with a preferred larger critical energy density around the matter-radiation equality time or an earlier recombination of protons and electrons before the photon decoupling can lead to an alleviation of the $H_0$ tension, since they allow for lower values of the comoving sound horizon at the baryon drag epoch, $r_d$, what in turn gives room for larger values of the Hubble parameter, which is needed to keep the good description of the BAO data and the location of the first peak of the CMB temperature anisotropies. This is the case of some models that have been proposed to alleviate the $H_0$ tension, based on early dark energy \citep{Poulin:2018cxd,Niedermann:2019olb,Agrawal:2019lmo,Gomez-Valent:2021cbe}, modified gravity \citep{SolaPeracaula:2019zsl,SolaPeracaula:2020vpg,Braglia:2020iik,Braglia:2020auw}, primordial magnetic fields \citep{Jedamzik:2020krr}, varying atomic constants \cite{Liu:2019awo,Sekiguchi:2020teg}, and running vacuum models \citep{SolaPeracaula:2021gxi}.

The measurement of $r_d$ has been carried out using a plethora of techniques in the past:

\begin{itemize}
\item From CMB data alone or in combination with other data sets, assuming concrete cosmological models. See e.g. \citep{Verde:2016wmz,Planck:2018vyg,Benisty:2020otr} and the references of the previous paragraph. 
\item Using the local (distance ladder) $H_0$ value as an anchor at $z=0$, together with BAO and SNIa data, assuming the $\Lambda$CDM \citep{Cuesta:2014asa} or using cubic splines to reconstruct the Hubble function \citep{Bernal:2016gxb,Aylor:2018drw}. Alternatively, it is also possible to use the local calibration of the absolute magnitude of SNIa from SH0ES, again with BAO and SNIa data \cite{Camarena:2019rmj}.
\item Fixing the scale with cosmic chronometers (CCH), employed in combination with BAO and SNIa, either reconstructing the shape of $H(z)$ with linearly-interpolated values that are left free in the fit \citep{Heavens:2014rja,Verde:2016ccp}, with a Multi-Task Gaussian Process \citep{Haridasu:2018gqm}, or using alternative parametrizations \cite{Cai:2021weh}.
\item Model-independent determinations of the product $H_0r_d$ are also possible using BAO and SNIa data alone \cite{LHuillier:2016mtc,Shafieloo:2018gin,Camarena:2019rmj}.
\end{itemize}
                                                          
We propose here a new method that lets us measure $r_d$ and $M$ under very mild theoretical assumptions (which reduce mainly to the validity of the Cosmological Principle and the metric description of gravity), and without making use of the two data sets driving the $H_0$ tension, namely the CMB and the calibration of the SNIa performed in the first steps of the distance ladder by SH0ES. Our method is close in spirit to those employed in \cite{Heavens:2014rja,Verde:2016ccp,Haridasu:2018gqm}, since we also use data on BAO, SNIa and CCH, and do not assume a particular cosmological model. Here we take, though, a different approach. While in \cite{Heavens:2014rja,Verde:2016ccp} the authors used a spline for $H(z)$ and reconstructed it, allowing the values of $H(z_i)$ at several redshift nodes $z_i$ to vary in the fitting analysis (together with $M$, $r_d$ and the curvature density parameter $\Omega^0_k$), here we only vary the last three. Our method is also different from the one presented in \citep{Haridasu:2018gqm,Cai:2021weh}. In this work we do not aim to find the shape of $H(z)$ preferred by the low-redshift data. We measure the size of the cosmic ruler and the SNIa intrinsic brightness by minimizing a loss function that tells us what is the level of inconsistency between the BAO, SNIa and CCH data sets for each triplet of values $\vec{\theta}=(M,r_d,H_0^2\Omega^0_k)$. Every $\vec{\theta}$ can be used to translate the BAO+SNIa+CCH data into a list of cosmological distances and values of the Hubble rate at different redshifts, with their corresponding covariance matrix. The difference between the distances and Hubble rates inferred from these data sets is clearly a function of $\vec{\theta}$ and, in particular, is very sensitive to the calibrators $M$ and $r_d$. In absence of unaccounted systematic errors in the data we should expect the calibrators $M$ and $r_d$ to minimize these differences. Thus, we can associate a larger probability to a vector $\vec{\theta}$ that leads to a lower statistical tension between data sets. In this work we employ the index of inconsistency (IOI) proposed in \cite{Lin:2017ikq} to quantify the level of disagreement between the transformed data sets and obtain the posterior distribution for the three parameters contained in $\vec{\theta}$. The IOI has been previously applied in \cite{Lin:2017ikq,Lin:2017bhs,Garcia-Quintero:2019cgt,Gonzalez:2021ojp} to estimate the tension between Gaussian posterior distributions of parameters that are obtained from the fitting analyses of the same model, but using alternative data sets. Hence, it was used as a measure of the existing tensions between data sets in the context of concrete models, e.g. the $\Lambda$CDM. Here we deal only with data, i.e. no cosmological model nor parametrization is assumed, with the aim to reduce as much as we can the degree of model dependence of our results. We demand the minimization of the inconsistency between the SNIa, BAO and CCH data sets through the optimal choice of the calibrators contained in $\vec{\theta}$. Our method has not been employed before in the literature. Model-independent approaches as the one proposed and studied in this paper will allow us to further test the viability of the candidate models aiming to loosen the $H_0$ tension, as well as the calibration of the SNIa in the first steps of the distance ladder.

The paper is organized as follows. In Sec. \ref{sec:data} we describe the data sets that we have employed in our analysis. In Sec. \ref{sec:method} we explain our statistical method. Our results are presented in Sec. \ref{sec:results}, and the conclusions in Sec. \ref{sec:conclusions}. 


\section{Data sets and observables}\label{sec:data}

These are the data sets that we have employed in our study.


\subsection{Supernovae of Type Ia}\label{sec:SNIa}

The expression for the apparent magnitude $m(z)$ of standardized SNIa reads,

\begin{equation}
m(z) = M +25 +5\log_{10}\left(\frac{D_L(z)}{1\,{\rm Mpc}}\right)\,,
\end{equation}
with $M$ the absolute magnitude and $D_L(z)$ the luminosity distance. In a Friedmann-Lema\^itre-Robertson-Walker universe the latter takes the following form,

\begin{equation}\label{eq:lumDista}
D_L(z) = \frac{c(1+z)}{\sqrt{\Omega_k^{0}H_0^2}}\sinh\left(\sqrt{\Omega_k^0H_0^2}\int_0^{z}\frac{dz^\prime}{H(z^\prime)}\right)\,,
\end{equation}
where $\Omega_k^0=-kc^2/(R_0H_0)^2$ is the curvature density parameter, with $k=0,-1,+1$ for a flat, open and closed universes, respectively. $R_0$ is a constant with units of length that can be interpreted as the current radius of curvature in a closed universe.

In this work we employ the SNIa contained in the Pantheon compilation \citep{Scolnic:2017caz}. For practical purposes we opt to use the 40 binned data points provided in \footnote{http://github.com/dscolnic/Pantheon}, with their corresponding covariance matrix, which includes the statistical and systematic uncertainties. The bins span the redshift range $z\in [0.014,1.6123]$. 


\subsection{Baryon acoustic oscillations}\label{sec:BAO}

Baryons and photons were tightly coupled through electromagnetic interactions during the pre-recombination era. The intense fight between radiation pressure and gravity in the photon-baryon plasma generated sound waves that left an imprint in the distribution of baryons when the universe was cool enough for the CMB photons to escape and start their travel towards us. The maximum distance traveled by this wave, the sound horizon at the baryon drag epoch, is prompted in the distribution of matter in the universe. It manifests as a peak in the matter two-point correlation function or as wiggles in the matter power spectrum \citep{Cole:2005sx,Eisenstein:2005su}. Galaxy surveys use this characteristic length, $r_d$, as a standard ruler with respect to which they can measure cosmological distances at various redshifts. The latter can be employed to constrain cosmological models in a quite robust way \citep{Bernal:2020vbb}. Their constraints are given either in terms of the dilation scale $D_V$,
\begin{equation}
\frac{D_V(z)}{r_d}=\frac{1}{r_d}\left[D_M^2(z)\frac{cz}{H(z)}\right]^{1/3}\,,
\end{equation}
with $D_M=(1+z)D_{A}(z)$ being the comoving angular diameter distance, or by splitting (when possible) the transverse and line-of-sight BAO information, providing data on $D_{A}(z)/r_d$ and $H(z)r_d$ separately, with some degree of correlation. In any metric theory of gravity with photons traveling on null geodesics and conservation of the photon number, the Etherington relation \citep{Etherington:1933} applies, 

\begin{equation}\label{eq:AngDista}
D_A(z)=\frac{D_L(z)}{(1+z)^2}\,
\end{equation}
with $D_{L}(z)$ given by Eq. (\ref{eq:lumDista}). This expression is useful to translate luminosity distances into angular diameter distances, and viceversa. 

We employ the following BAO data points:

\begin{itemize}

\item $D_V/r_d$ at $z=0.122$ provided in \citep{Carter:2018vce}, which combines the dilation scales previously reported by the 6dF Galaxy Survey (6dFGS) \citep{Beutler:2011hx} at $z=0.106$ and the one obtained from the Sloan Digital Sky Survey (SDSS) Main Galaxy Sample at $z=0.15$ \citep{Ross:2014qpa}.

\item The anisotropic BAO data measured by BOSS using the LOWZ ($z=0.32$) and CMASS ($z=0.57$) galaxy samples \citep{Gil-Marin:2016wya}.

\item The dilation scale measurements by WiggleZ at $z=0.44,0.60,0.73$ \citep{Kazin:2014qga}. 

\item $D_A/r_d$ at $z=0.81$ measured by the Dark Energy Survey Year 1 (DESY1) \citep{DES:2017rfo}. We will also study the impact of substituting this point by the more recent measurement from DESY3 at the effective redshift $z=0.835$ \citep{DES:2021esc}, which is in $2.3\sigma$ tension with the {\it Planck} prediction assuming the $\Lambda$CDM.

\item  The anisotropic BAO data from the extended BOSS Data Release 16 (DR16) quasar sample at $z=1.48$ \citep{Neveux:2020voa,Hou:2020rse}.

\end{itemize}

We avoid the use of the anisotropic BAO data obtained from the Ly$\alpha$ absorption and quasars of the final data release (SDSS DR16) of eBOSS at $z=2.334$ \citep{duMasdesBourboux:2020pck} because it falls out of the measurement ranges of SNIa and CCH, see Sec. \ref{sec:SNIa} and \ref{sec:CCH}. The full BAO data vector and associated covariance matrix is provided in Appendix \ref{sec:appA}.


\subsection{Cosmic chronometers}\label{sec:CCH}

Spectroscopic dating techniques of passively evolving galaxies, i.e. galaxies with old stellar populations and low star formation rates, have become a good tool to obtain observational values of the Hubble function at redshifts $z\lesssim 2$ \citep{Jimenez:2001gg}. These measurements do not rely on any particular cosmological model, although are subject to other sources of systematic uncertainties, as to the ones associated with the modeling of stellar ages, see e.g. \cite{Moresco:2020fbm}, which is carried out through the so-called stellar population synthesis (SPS) techniques, and also with a possible contamination due to the presence of young stellar components in quiescent galaxies \citep{Lopez-Corredoira:2017zfl,Lopez-Corredoira:2018tmn,Moresco:2018xdr}. Given a pair of ensembles of passively evolving galaxies at two different redshifts it is possible to infer $dz/dt$ from observations under the assumption of a concrete SPS model and compute $H(z) = -(1 + z)^{-1}dz/dt$. Thus, cosmic chronometers allow us to obtain the value of the Hubble function at different redshifts, contrary to other probes which do not directly measure $H(z)$, but integrated quantities as e.g. luminosity distances.

In this study we use the 24 data points on $H(z)$ from CCH in the redshift range $z\in [0.07,1.965]$ reported in \citep{Jimenez:2003iv,Stern:2009ep,Moresco:2012jh,Zhang:2012mp,Moresco:2015cya,Moresco:2016mzx,Ratsimbazafy:2017vga,Borghi:2021rft}. See also Fig. \ref{fig:Hrec} in Appendix \ref{sec:appC}. More concretely, in our main analyses we make use of the {\it processed} sample provided in Table 2 of \citep{Gomez-Valent:2018gvm}, but adding the data point of \citep{Borghi:2021rft} and removing the ones of \citep{Simon:2004tf}\footnote{Serious concerns about the statistical analysis carried out in \citep{Simon:2004tf} have been recently raised in \cite{Kjerrgren:2021zuo}. Thus, we prefer to omit at the moment the use of these CCH data.}. Our resulting CCH data set is robust, since it introduces corrections accounting for the systematic errors mentioned above. In addition, we also study the effect of the covariance matrix of the CCH data given in \cite{Moresco:2020fbm}. We compute it with the code \footnote{https://gitlab.com/mmoresco/CCcovariance}, cf. Sec. \ref{sec:results} and the comments in Appendix \ref{sec:appC}.


\section{The method}\label{sec:method}

If the SNIa, BAO and CCH data sets described in the previous section are unbiased they should be, of course, consistent with each other. The level of consistency, though, depends on the values of the parameters contained in the vector $\vec{\theta}=(M,r_d,H_0^2\Omega_k^0)$, since they are used to translate the original BAO and SNIa data into a set of cosmological distances and Hubble rates. Our goal in this paper is to infer the distribution of these parameters by maximizing the consistency of the data sets under consideration. To do so we need first to find or define a function $L(\vec{\theta})$ that lets us quantify the degree of inconsistency between the data sets for every $\vec{\theta}$, and then use it to build a distribution that maps the latter to concrete probability values. The function $L(\vec{\theta})$ should include the contribution of the inconsistency estimates of all pairs of data sets. $L(\vec{\theta})$ can be thought of as a loss function, since for wide enough priors the distribution that we build from it will be maximized for the value of $\vec{\theta}$ that minimizes $L$. We construct it making use of the index of inconsistency proposed in \cite{Lin:2017ikq}, as follows,

\begin{equation}\label{eq:L}
L(\vec{\theta}) = L_1(M,r_d)+L_2(M,H_0^2\Omega_k^0)\,,
\end{equation}
with 

\begin{equation}\label{eq:L1}
L_1(M,r_d)={\rm IOI}[{\rm BAO,SNIa+CCH}]
\end{equation} 
and
\begin{equation}\label{eq:L2}
L_2(M,H_0^2\Omega_k^0)={\rm IOI}[{\rm SNIa,CCH}]\,.
\end{equation} 
\eqref{eq:L1} is the index of inconsistency between the BAO data and the string SNIa+CCH,  which accounts for the inconsistencies between the pairs (BAO,SNIa) and (BAO,CCH). On the other hand, \eqref{eq:L2} is the IOI between the SNIa and the CCH data sets. The two-experiment IOI is defined as follows \citep{Lin:2017ikq},

\begin{equation}\label{eq:IOI}
{\rm IOI}[i,j]=\frac{1}{2}\delta^{T}(C^{(i)}+C^{(j)})^{-1}\delta\,,
\end{equation} 
where $C^{(i)}$ is the covariance matrix of the ith data set and $\delta=\mu^{(i)}-\mu^{(j)}$ is the difference between the data vectors of the two data sets under consideration, i.e. data sets $i$ and $j$. The IOI is a generalization of the Mahalanobis distance \citep{Mahalanobis:1936}, and strictly speaking it is reliable only for Gaussian-distributed data sets\footnote{Fortunately, this is the case in the current study, in very good approximation. We will comment on this explicitly when needed.}. The specific dependence of $L_1$ and $L_2$ on the parameters of $\vec{\theta}$ and all the details about how to build them will be duly explained below,  in Secs. \ref{sec:IOI1} and \ref{sec:IOI2}.

Now the question is how to build the probability distribution out of \eqref{eq:L}. We consider the following form,

\begin{equation}\label{eq:dist}
P(\vec{\theta})=\pi(\vec{\theta})\times\mathcal{N}e^{-L(\vec{\theta})}\,,
\end{equation}   
where $\pi(\vec{\theta})$ is the prior distribution of the parameters and $\mathcal{N}$ is a constant normalization factor. The motivation for the exponential term is that given a pair of calibrated data sets $i=1$ and $j=2$ the value of IOI$[1,2]$ is exactly the same as the one that would be obtained by considering a fake theoretical point located at $\mu^{(1)}$ and fake data centered at $\mu^{(2)}$ with covariance matrix $C^{(1)}+C^{(2)}$. Different values of $\vec{\theta}$ lead to a different pair of calibrated data sets, of course, but their associated IOI can be mapped to the one obtained from a different fake theoretical point with exactly the same fake data as considered before. Let us express this in more mathematical terms. Given a reference fake data vector $\tilde{\mu}$ with covariance matrix $\tilde{C}$ it is always possible to perform a change of coordinates such that

\begin{equation}
(C^{(i)}+C^{(j)}) = B^{\rm T}\tilde{C}B\,,
\end{equation}
with $B$ the transformation matrix associated to the aforesaid change of coordinates, which will obviously depend on the parameters entering the covariance matrices $C^{(i)}$ and $C^{(j)}$. Using this we can do

\begin{align}
(\mu^{(i)}-\mu^{(j)})^{\rm T}&(C^{(i)}+C^{(j)})(\mu^{(i)}-\mu^{(j)})= \\
&(\mu^{(i)}-\mu^{(j)})^{\rm T}B^{\rm T}\tilde{C}B(\mu^{(i)}-\mu^{(j)})\,.
\end{align}
Thus, if we define

\begin{equation}
\mu^{(i,j)} -\tilde{\mu}\equiv B(\mu^{(i)}-\mu^{(j)})\,,
\end{equation}
we can rewrite \eqref{eq:IOI} as follows,

\begin{equation}\label{eq:IOI2}
{\rm IOI}[i,j] = \frac{1}{2}(\mu^{(i,j)}-\tilde{\mu})^{\rm T}\tilde{C}(\mu^{(i,j)}-\tilde{\mu})\,.
\end{equation} 
One can map every set of data $(i,j)$ to a particular fake theoretical point $\mu^{(i,j)}$, considering a data vector $\tilde{\mu}$ with covariance matrix $\tilde{C}$. The only element that depends on $\vec{\theta}$ in the last formula is the fake theoretical point $\mu^{(i,j)}$. It is then natural to consider \eqref{eq:dist}, as in standard $\chi^2$ analyses. Moreover, if data are unbiased and our minimal theoretical assumptions hold \eqref{eq:dist} leads to the correct estimation of the calibrators, with their uncertainties decreasing for an increasing number of data points. This can be easily understood. Imagine that we add unbiased data to our problem, let us say with a similar constraining power as the preceding data set. If we compute the ratios between the probability density at the peak $\hat{\theta}$ and at a point $\tilde{\theta}$ away from it before and after adding the new data we find that these ratios read: $R_{\rm old}=e^{L(\tilde{\theta})-L(\hat{\theta})}$, $R_{\rm new}\sim e^{2(L(\tilde{\theta})-L(\hat{\theta}))}$, so $R_{\rm new}/R_{\rm old}\sim R_{\rm old}>1$. This means that after the addition of the new data the point in parameter space located far away from the peak is less preferred than before when compared to the one at which \eqref{eq:dist} is maximized, what in turn leads to lower uncertainties for the parameters contained in $\vec{\theta}$. 

Once we have the probability distribution \eqref{eq:dist} we can compute the confidence regions for the parameters of $\vec{\theta}$ either by evaluating our distribution on a three-dimensional grid or by sampling it with a Monte Carlo method. We opt to do the latter, see the details in Sec. \ref{sec:MC}.

\begin{figure*}[t!]
\begin{center}
\includegraphics[width=6.5in, height=3.5in]{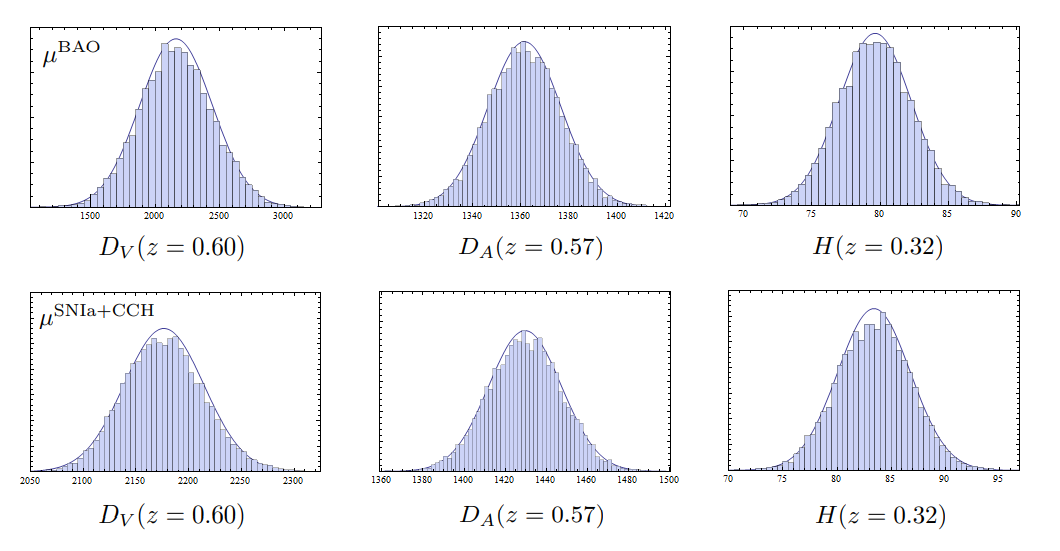}
\caption{This figure shows the histograms obtained for three out of the eleven elements of $\mu^{\rm BAO}$ (first row) and $\mu^{\rm SNIa+CCH}$ (second row), obtained using $10^4$ points in the sampling process and using $M=-19.454$ and $r_d=145$ Mpc. The values of the dilation scale $D_V(z=0.60)$ and the angular diameter distance $D_A(z=0.57)$ are expressed in Mpc, whereas the Hubble function $H(z=0.32)$ is in km/s/Mpc. As mentioned in the main text, the distributions are very well approximated by Gaussians with the mean and standard deviation inferred from the sample. The latter are also plotted with solid lines, in blue. See Sec. \ref{sec:IOI1} for details.}\label{fig:histo_IOI1}
\end{center}
\end{figure*}


\subsection{IOI between BAO and SNIa+CCH}\label{sec:IOI1}

Given a value of $r_d$ we can rewrite the BAO constraints on $\{D_V(z)/r_d,D_A(z)/r_d,H(z)r_d\}$ of Sec. \ref{sec:BAO} (cf. \eqref{eq:BAOdata}-\eqref{eq:BAOcov} in Appendix \ref{sec:appA}) as constraints on $\{D_V(z),D_A(z),H(z)\}$, with a data vector

\begin{equation}
\mu^{\rm BAO}(r_d) = \begin{pmatrix}
D_V(z=0.122)\\
D_V(z=0.44)\\
D_V(z=0.60)\\
D_V(z=0.73)\\
D_A(z=0.32)\\
D_A(z=0.57)\\
D_A(z=0.81)\\
D_A(z=1.48)\\
H(z=0.32)\\
H(z=0.57)\\
H(z=1.48)
\end{pmatrix}\,,
\end{equation}
and its associated covariance matrix $C^{\rm BAO}(r_d)$. Both depend on $r_d$. The resulting distribution is Gaussian in very good approximation, as we have explicitly checked for different values of the sound horizon, cf. Fig. \ref{fig:histo_IOI1}. In order to compute the IOI between the BAO and SNIa+CCH data sets we need to build the analogous quantities from the SNIa+CCH joint data set, i.e. $\mu^{\rm SNIa+CCH}$ and $C^{\rm SNIa+CCH}$. As we do not have SNIa and CCH data at the exact BAO redshifts, we need first to compute the extrapolated values of $D_A(z)$ and $H(z)$ at these $z$'s using some technique that allows us to get the vectors

 \begin{equation}\label{eq:inter}
\mu^{\rm SNIa}(M) = \begin{pmatrix}
D_A(z=0.122)\\
D_A(z=0.44)\\
D_A(z=0.60)\\
D_A(z=0.73)\\
D_A(z=0.32)\\
D_A(z=0.57)\\
D_A(z=0.81)\\
D_A(z=1.48)
\end{pmatrix}
\end{equation}
and

\begin{equation}\label{eq:muCCH}
\mu^{\rm CCH}= \begin{pmatrix}
H(z=0.122)\\
H(z=0.44)\\
H(z=0.60)\\
H(z=0.73)\\
H(z=0.32)\\
H(z=0.57)\\
H(z=1.48)
\end{pmatrix}\,,
\end{equation}
with their individual covariances, $C^{\rm SNIa}(M)$ and $C^{\rm CCH}$. Notice that $\mu^{\rm SNIa}$ and $C^{\rm SNIa}$ depend on $M$, since we can only get angular diameter distances from the SNIa data after fixing the absolute magnitude of the supernovae. Using all these ingredients we can build the final mean and covariance matrix $\mu^{\rm SNIa+CCH}$ and $C^{\rm SNIa+CCH}$ to be employed (together with $\mu^{\rm BAO}$ and $C^{\rm BAO}$) in the computation of $L_1$ \eqref{eq:L1}.

For a given value of $M$, and making use of Eqs. \eqref{eq:lumDista} and \eqref{eq:AngDista}, we can translate the SNIa apparent magnitudes $m(z)$ into data on $D_A(z)$ at the redshifts of the (binned) SNIa data. By sampling this distribution and using e.g. a cubic interpolation method, it is easy to infer the Gaussian distribution for the $D_A$'s at the redshifts of interest, i.e. those specified in \eqref{eq:inter}. In order to be more efficient, though, it is better actually to split $D_A$ as follows, 

\begin{equation}\label{eq:Bfunc}
D_A(z)=10^{-M/5}B(z)\,{\rm Mpc}\quad{\rm with}\quad B(z)=\frac{10^{\frac{m(z)-25}{5}}}{(1+z)^2}\,,
\end{equation}
and sample the distribution of $m$'s to generate the mean ($\mu_B$) and covariance matrix ($C_B$) for the $B$'s at the redshifts of \eqref{eq:inter} before starting the Monte Carlo, since this part of $D_A$ does not depend on $M$ and hence can be employed at each step of the sampling process. This distribution is Gaussian too, see Appendix \ref{sec:appB}. Given a value of $M$, it is easy to obtain then the mean and covariance for $D_A$ from the distribution of $B(z)$. We just need to do: 

\begin{equation}
\mu^{\rm SNIa}= 10^{-M/5}\mu_B\quad;\quad C^{\rm SNIa}=10^{-2M/5}C_B\,.
\end{equation}
We proceed in a similar way to obtain $\mu^{\rm CCH}$ \eqref{eq:muCCH} and the corresponding covariance matrix $C^{\rm CCH}$. In this case the result does not depend on any of the parameters contained in $\vec{\theta}$. Therefore, we can employ the same vector $\mu^{\rm CCH}$ and covariance $C^{\rm CCH}$ in each step of the Monte Carlo routine. We reconstruct the shape of $H(z)$ using the data described in Sec. \ref{sec:CCH} and Gaussian Processes (GP), with a Gaussian kernel. In appendix \ref{sec:appC} we describe the GP reconstruction of the Hubble function and provide the resulting $\mu^{\rm CCH}$ and $C^{\rm CCH}$, and also comment on several technical aspects for the interested reader.

The obtention of $\mu^{\rm SNIa+CCH}(M)$ and $C^{\rm SNIa+CCH}(M)$ from ($\mu^{\rm SNIa}(M),C^{\rm SNIa}$(M)) and ($\mu^{\rm CCH},C^{\rm CCH}$) is straightforward. It can be done through a simple sampling of the two multivariate Gaussians. The result is in very good approximation Gaussian too, cf. again Fig. \ref{fig:histo_IOI1}. Equipped with these tools we can finally evaluate $L_1$ \eqref{eq:L1}, which is of course a function of the calibrators $M$ and $r_d$.


\begin{table*}[t!]
\centering
\begin{tabular}{|c  ||c | c |  c | c |   }
 \multicolumn{1}{c}{} & \multicolumn{1}{c}{} & \multicolumn{1}{c}{} & \multicolumn{1}{c}{}
\\\hline
{\small Loss function} & {\small $M$}  & {\small $r_d$ [Mpc]} & {\small $H_0^2\Omega_k^0$ [(km/s/Mpc)$^2$]}  &  {\small $\Omega_k^0$}
\\\hline
$L_1(M,r_d)$ & $-19.355\pm 0.095$ & $145.5^{+5.4}_{-6.0}$ & - & - \\\hline
$L(M,r_d,0)$, flat universe  & $-19.374\pm 0.080$ & $146.5^{+4.8}_{-5.7}$ & 0 & 0 \\\hline
$L(M,r_d,H_0^2\Omega_k^0)$ & $-19.362^{+0.078}_{-0.067}$ & $146.0^{+4.2}_{-5.1}$ & $-36\pm 490$ & $-0.01\pm 0.10 $ \\\hline
$\mathcal{L}(M,r_d,H_0^2\Omega_k^0)$ & $-19.363\pm 0.076$ & $146.7\pm 5.1$  & $-24\pm 490$ & $0.00\pm 0.10$\\\hline
$L_c(M,r_d,H_0^2\Omega_k^0)$ & $-19.399\pm 0.098$  & $148.3\pm 6.6$  & $-38\pm 490$  & $-0.01\pm 0.10$ \\\hline
\end{tabular}
\label{tab:table}
\caption{Means and uncertainties at the $1\sigma$ c.l. of the fitting parameters obtained with the loss functions listed in Sec. \ref{sec:MC}. The constraints on $\Omega^0_k$ are computed by breaking the degeneracy in the $H_0-\Omega_k^0$ plane with the Gaussian prior $H_0=(70.72\pm 6.44)$ km/s/Mpc (or $70.36\pm 5.69$ km/s/Mpc when we consider the covariance matrix of the CCH data) obtained from our reconstruction with GP (cf. Appendix \ref{sec:appC}). See the comments in Sec. \ref{sec:results} and Fig. \ref{fig:contours}.}
\end{table*}

\subsection{IOI between SNIa and CCH}\label{sec:IOI2}

We compute the index of inconsistency between the SNIa and CCH data sets by first noting that for a given pair $(M,H_0^2\Omega_k^0)$ we can translate a particular value of $m(z_i)$ into a value of the Hubble function $H(z_i)$, if we also know the derivative of the apparent magnitude at that redshift. This becomes evident if we perform the derivative of Eq. \eqref{eq:lumDista} with respect to the redshift, and isolate $H(z)$,  

\begin{equation}\label{eq:Hdeco}
H(z)=\frac{\left[\left(\frac{c(1+z)}{D_L(z)}\right)^2+H_0^2\Omega_k^0\right]^{1/2}}{\frac{\ln(10)}{5}\frac{\partial m}{\partial z}-\frac{1}{1+z}}\,, 
\end{equation}
where the luminosity distance can be written in terms of a function $\tilde{B}(z)$ that does not depend on $M$,

\begin{equation}\label{eq:Btildefunc}
D_L(z)=10^{-\frac{(M+25)}{5}}\tilde{B}(z)\,{\rm Mpc}\quad{\rm with}\quad \tilde{B}(z)=10^{\frac{m(z)}{5}}\,.
\end{equation} 
Formula \eqref{eq:Hdeco} is very useful to translate the SNIa data into data on the Hubble rate and establish the link with the CCH data. We can sample the distribution of $m$'s at the SNIa redshifts and obtain from it the distribution of $\tilde{B}$'s and $\partial m/\partial z$'s at those redshifts at which we have CCH data\footnote{We exclude the CCH data point at $z=1.965$ \cite{Moresco:2015cya} because we have no SNIa data at this high redshift.}, e.g. making use again of a cubic interpolation method. The resulting distribution is Gaussian. The derivative of the apparent magnitude can be computed numerically, using finite differences with $\Delta z=0.01$. Finally, given a pair $(M,H_0^2\Omega_k^0)$ we can construct from the latter the distribution of values of the Hubble function at the CCH redshifts using \eqref{eq:Hdeco}. It is obviously a function of $M$ and $H_0^2\Omega_k^0$. The computation of $L_2$ \eqref{eq:L2} is at this stage very easy because we already have all the ingredients to apply \eqref{eq:IOI}. In order to sample physically motivated values of the product $H_0^2\Omega_k^0$ we use the Gaussian prior

\begin{equation}\label{eq:prior}
\pi(H_0^2\Omega_k^0) = \frac{1}{\sqrt{2\pi}\sigma}e^{-\frac{(H_0^2\Omega_k^0)^2}{2\sigma^2}}\,,
\end{equation}
with $\sigma=500$ (km/s/Mpc)$^2$. The latter is motivated by the constraint on $H_0$ obtained from our GP reconstruction, $H_0=(70.72\pm 6.44)$ km/s/Mpc (cf. Appendix \ref{sec:appC}), and also by the ones on $\Omega_k^0$ obtained under some CMB data sets \citep{Planck:2018vyg,Handley:2019tkm,DiValentino:2019qzk}, which allow values of $\Omega_k^0\sim -0.1$ at $\sim 1.5\sigma$ c.l. Much tighter constraints on the curvature parameter are derived when CMB lensing, SNIa and/or BAO data are also considered \citep{Planck:2018vyg,Efstathiou:2020wem}, but we want to proceed as model-independently as possible. This is why we choose a wide prior for this parameter, but still forbidding values that clearly fall out of the region allowed by the CMB. We center the Gaussian prior at $\Omega_k^{0}=0$. In this way we keep a reasonable upper bound on the absolute value of this parameter, without favoring a particular sign. As we will discuss in Sec. \ref{sec:results}, $H_0^2\Omega_k^0$ has a very low impact on our constraints on $M$ and $r_d$, and its posterior is basically dominated by the prior.


\subsection{Monte Carlo analyses}\label{sec:MC}

We sample the distributions \eqref{eq:dist} built with the following functions: 

\begin{itemize}
\item $L_1(M,r_d)$, Eq. \eqref{eq:L1}.
\item $L(M,r_d,H_0^2\Omega_k^0=0)$ assuming a flat universe. 
\item $L(M,r_d,H_0^2\Omega_k^0)$.
\item $L(M,r_d,H_0^2\Omega_k^0)$, but using the data point from DESY3 \citep{DES:2021esc} instead of DESY1 \citep{DES:2017rfo}. We call this loss function $\mathcal{L}(M,r_d,H_0^2\Omega_k^0)$.
\item $L(M,r_d,H_0^2\Omega_k^0)$, but considering the covariance matrix of the CCH data \cite{Moresco:2020fbm}. We denote this loss function $L_c(M,r_d,H_0^2\Omega_k^0)$.
\end{itemize}
We make use of the Metropolis–Hastings algorithm \citep{Metropolis:1953,Hastings:1970}. For the calibrators we use the flat priors $M\in [-19.7,-19.0]$ and $r_d\in[110,170]$ Mpc, which are much wider than their associated uncertainties and hence have no impact on their posterior distributions. The results obtained in our Monte Carlo analyses are presented and discussed in the next section.


\begin{figure*}[t!]
\begin{center}
\includegraphics[width=5in, height=4.5in]{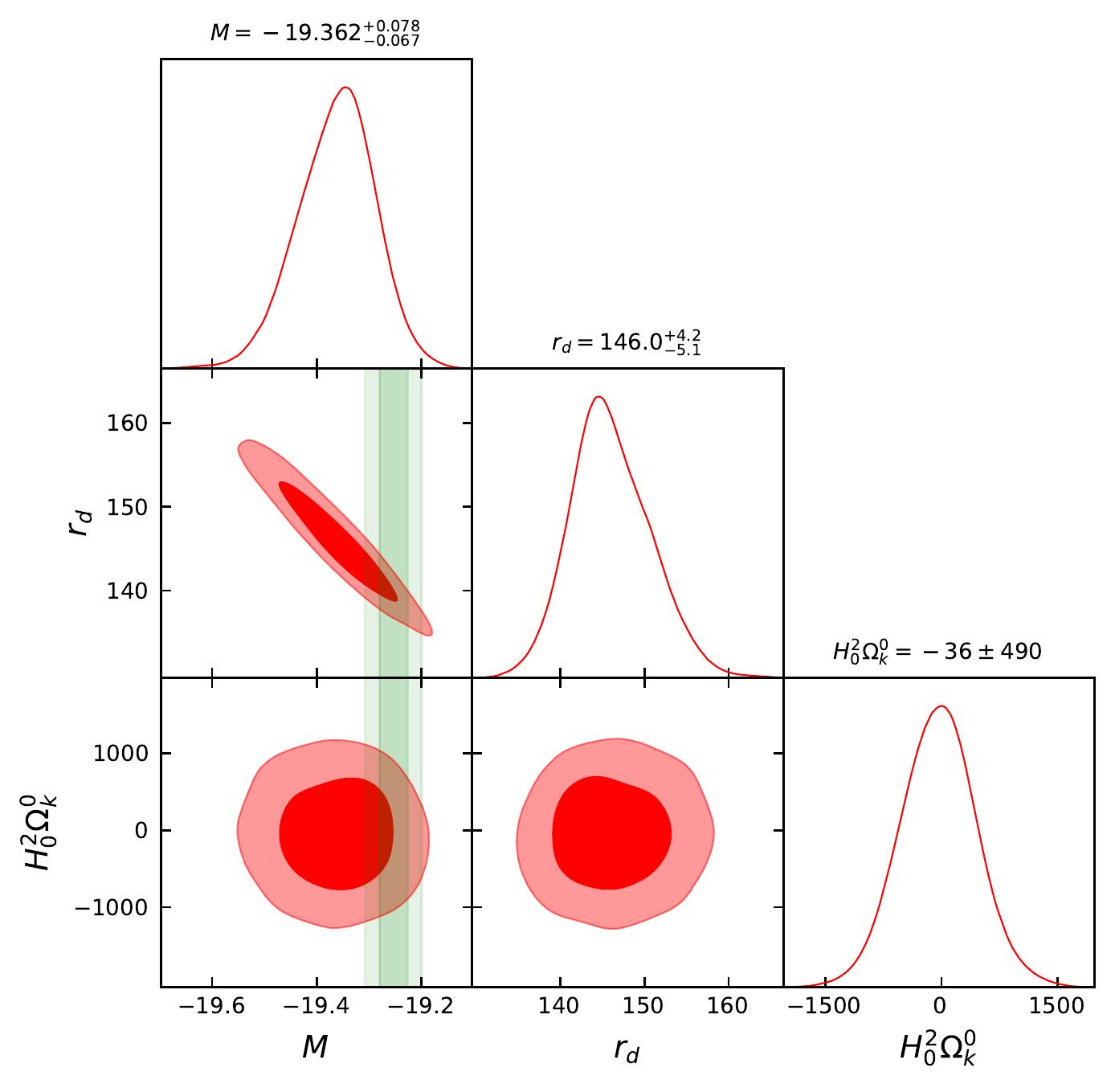}
\caption{Contour plots at $1\sigma$ and $2\sigma$ c.l. in all the planes of the ($M,r_d,H_0^2\Omega_k^0$)-parameter space, obtained with the loss function $L(M,r_d,H_0^2\Omega_k^0)$, cf. Sec. \ref{sec:MC}. $r_d$ is expressed in Mpc and $H_0^2\Omega_k^0$ in (km/s/Mpc)$^2$. The vertical green bands correspond to the (1$\sigma$ and 2$\sigma$) constraint on the absolute magnitude of SNIa obtained by SH0ES from the first steps of the distance ladder \cite{Riess:2021jrx}, $M=-19.253\pm0.027$. Our constraint on $H_0^2\Omega_k^0$ is basically dominated by the prior \eqref{eq:prior}. In the absence of this prior we would obtain much more elongated contours in the $y$-direction both in the $(M,H_0^2\Omega_k^0)$ and $(r_d,H_0^2\Omega_k^0)$ planes. We have made use of \texttt{GetDist} \cite{Lewis:2019xzd} to analyze the Markov chains and generate this figure. See Sec. \ref{sec:results} for further comments.}\label{fig:contours}
\end{center}
\end{figure*}

\section{Results}\label{sec:results}

Our constraints on $M$, $r_d$, $H_0^2\Omega^0_k$ and $\Omega^0_k$ are shown in Table I. In Fig. \ref{fig:contours} we plot the confidence regions at $1$ and $2\sigma$ c.l. in all the relevant planes of the parameter space, together with the corresponding one-dimensional posteriors. The central values for $M$ and $r_d$ are compatible in all the scenarios analyzed in this study, even when the loss function is only built with the IOI(BAO,SNIa+CCH), i.e. when $L=L_1$. In this case we find $M=-19.355\pm 0.095$ and $r_d=(145.5^{+5.4}_{-6.0})$ Mpc. If we penalize also the degree of inconsistency between SNIa and CCH by considering the loss function $L=L_1+L_2$ we find, though, that the uncertainties on $M$ and $r_d$ decrease by $\sim 20\%$, yielding $M=-19.362^{+0.078}_{-0.067}$ and $r_d=(146.0^{+4.2}_{-5.1})$ Mpc. These are our main results. Notice that the inclusion of $L_2$ is important to study also the compatibility between the CCH and SNIa data sets, which is not automatically ensured with the use of $L_1$ alone. It allows us to put more stringent and more precise limits on the calibrators, i.e. to better draw the borders of the allowed region in parameter space. The absolute magnitude of SNIa is lower than the SH0ES measurement, at $\sim 1.4\sigma$ c.l. (cf. Fig. \ref{fig:contours}). This result is compatible with \cite{Efstathiou:2020wxn,Renzi:2020fnx}. Using the value of $M-5\log_{10}\left(H_0{\rm Mpc}/c\right)\sim -1.185$ obtained by SH0ES from the analysis of the SNIa in the Hubble flow we obtain the following values of $H_0$ associated to our $M-\sigma_M$, $M$ and $M+\sigma_M$, respectively: $H_0=(67.3,69.5,72.0)$ km/s/Mpc. The central value of the Hubble parameter is very close to the distance ladder measurement obtained with the TRGB calibration \cite{Freedman:2019jwv,Freedman:2021ahq}. However, it lies at $<1\sigma$ from the Planck result obtained under the standard model \cite{Planck:2018vyg} and at $\sim 1\sigma$ from the SH0ES value \citep{Riess:2021jrx}. This departure is not significant, but it will be interesting to revisit this calculation in the future to check whether this difference grows. If so, and if not caused by systematics in the data, it could indicate a flaw in our minimal (although still model-dependent) assumptions \cite{Macpherson:2021gbh,Krishnan:2021dyb}. Similarly, if the constraints on the calibrators $M$ and $r_d$ obtained with our method from different subsets of future CCH+BAO+SNIa data exhibit some level of tension this could be due, again, to systematic biases in the data or to some incorrect theoretical assumptions.

On the other hand, the comoving sound horizon agrees with the $\Lambda$CDM-based inference by {\it Planck}, although the one and two sigma bands also encompass much lower values of $r_d$, as those found e.g. in early dark energy and modified gravity scenarios that have been explored in the literature to alleviate (in greater or lesser extent) the $H_0$ tension. The correlation coefficient in the $M-r_d$ plane is negative, as expected, since larger values of $M$ lead also to larger values of the Hubble function and, therefore, $r_d$ needs to decrease in order to calibrate appropriately the BAO ruler. 

The error bars of both, $M$ and $r_d$, are a bit smaller if a flat universe is taken for granted, of course, just because we reduce by one the dimensionality of our parameter space. It is also worth to mention that the change of the DESY1 BAO data point \cite{DES:2017rfo} by the DESY3 one \cite{DES:2021esc}, which is in $2.3\sigma$ tension with the {\it Planck} $\Lambda$CDM best-fit cosmology, does not alter our results significantly. The impact of considering the covariance matrix of the CCH data from \cite{Moresco:2020fbm} is not important either, and our constraints on the curvature parameter are weak and essentially dominated by the prior (cf. again Table I). We find $\Omega_k^0=-0.01\pm 0.10$.


\section{Conclusions}\label{sec:conclusions}

The absolute magnitude of supernovae of Type Ia, $M$, and the comoving sound horizon, $r_d$, are the anchors of the direct and inverse cosmic distance ladders, respectively, and therefore play a very important role in the discussion of the $H_0$ tension. Many models aiming to solve the latter modify the physics of the pre-recombination era and predict a value of $r_d$ which is much lower than the one preferred by the $\Lambda$CDM when it is constrained with CMB data. In this paper we have employed low-$z$ data to constrain $M$ and $r_d$ under minimal model assumptions (which reduce to the Cosmological Principle and an underlying metric theory of gravity) and also avoiding the use of any input coming from the main drivers of the $H_0$ tension, similarly to what was done by other groups in the past \cite{Heavens:2014rja,Verde:2016ccp,Haridasu:2018gqm}. We have applied a novel method, though, based on the minimization of a loss function that quantifies the degree of inconsistency between the BAO, SNIa and cosmic chronometer data sets, which is built from the IOI estimator proposed by Lin and Ishak \cite{Lin:2017ikq}. It is only a function of $M$, $r_d$ and $H_0^2\Omega_k^0$. Our constraints read $M=-19.362^{+0.078}_{-0.066}$ and $r_d=(146.0^{+4.2}_{-5.1})$ Mpc at $1\sigma$ c.l. The former is slightly lower ($1.4\sigma$) than the value inferred in the first steps of the cosmic distance ladder by SH0ES, $M=-19.253\pm 0.027$ \cite{Riess:2021jrx}, whereas the comoving sound horizon is not very constrained by our principle of consistency. It is fully compatible with the standard model value from the TT,TE,EE+lowE+lensing analysis by {\it Planck} \cite{Planck:2018vyg}, $r_d=(147.09\pm 0.26)$ Mpc, but still leaves plenty of room for new physics \cite{Poulin:2018cxd,Niedermann:2019olb,Agrawal:2019lmo,Gomez-Valent:2021cbe,SolaPeracaula:2019zsl,SolaPeracaula:2020vpg,Braglia:2020iik,Braglia:2020auw,Jedamzik:2020krr,Liu:2019awo,Sekiguchi:2020teg,SolaPeracaula:2021gxi,Verde:2016wmz}. With the advent of future data e.g. from Euclid and LSST, our method will provide tighter constraints on these relevant parameters, allowing us to study the viability of cosmological models with non-standard pre-recombination physics and to test the agreement between the two ends of the cosmic ladder in a quite model-independent way.


\vspace{0.25cm}
\noindent {\bf Acknowledgements}
\newline
\newline
\noindent The author is funded by the Istituto Nazionale di Fisica Nucleare (INFN) through the project ``Dark Energy and Modified Gravity Models in the light of Low-Redshift Observations'' (n. 22425/2020). He is grateful to the Institute for Theoretical Physics (ITP) Heidelberg for letting him use its computational facilities from remote, and to Dr. Elmar Bittner for his valuable technical help in the use of the ITP computational resources. He would also like to thank Prof. Joan Sol\`a Peracaula for feedback on this manuscript, and Dr. Javier de Cruz P\'erez for pointing out the existence of the CCH covariance matrix. Finally, the author wants to express his gratitude to the anonymous referee for their insightful suggestions.


\appendix

\section{BAO data}\label{sec:appA}

These are the full data vector and covariance matrix of the BAO data described in Sec. \ref{sec:BAO}:

 \begin{equation}\label{eq:BAOdata}
 \vec{d}_{\rm BAO}=\begin{pmatrix}
D_V(z=0.122)/r_d\\
D_V(z=0.44)/r_d\\
D_V(z=0.60)/r_d\\
D_V(z=0.73)/r_d\\
D_A(z=0.32)/r_d\\
D_A(z=0.57)/r_d\\
D_A(z=0.81)/r_d\\
D_A(z=1.48)/r_d\\
r_dH(z=0.32)\\
r_dH(z=0.57)\\
c/(r_dH(z=1.48))
\end{pmatrix}=\begin{pmatrix}
3.65\\
11.55\\
14.95\\
16.93\\
6.60\\
9.39\\
10.75\\
12.18\\
11549\\
14021\\
13.23
\end{pmatrix}\,,
\end{equation}

\begin{widetext}
\begin{equation}\label{eq:BAOcov}
\mathcal{D}_{\rm BAO}=\begin{pmatrix}
0.0144 & 0 & 0 & 0 & 0 & 0 & 0 & 0 & 0 & 0 & 0 \\
- & 4.812 & -2.465 & 1.037 & 0 & 0 & 0 & 0 & 0 & 0 & 0 \\
- & - & 3.770 & -1.587 & 0 & 0 & 0 & 0 & 0 & 0 & 0 \\
- & - & - & 3.650 & 0 & 0 & 0 & 0 & 0 & 0 & 0 \\
- & - & - & - & 0.0179 & 0 & 0 & 0 & 28.929 & 0 & 0 \\
- & - & - & - & - & 0.0106 & 0 & 0 & 0 & 13.827 & 0 \\
- & - & - & - & - & - & 0.1849 & 0 & 0 & 0 & 0 \\
- & - & - & - & - & - & - & 0.101 & 0 & 0 & 0.0057 \\
- & - & - & - & - & - & - & - & 148099 & 0 & 0 \\
- & - & - & - & - & - & - & - & - & 50717 & 0 \\
- & - & - & - & - & - & - & - & - & - & 0.2195
\end{pmatrix}\,.
\end{equation}
\end{widetext}
The product $r_dH(z)$ is given in km/s. The other quantities are dimensionless. We do not specify the lower half of the covariance matrix (it is of course symmetric). When the data point from DESY3 \cite{DES:2021esc} is used instead of the one from DESY1 \cite{DES:2017rfo}, we change $D_A(z=0.81)/r_d=10.75\pm 0.43$ by $D_A(z=0.835)/r_d=10.31\pm 0.28$. 

\begin{figure*}[t!]
\begin{center}
\includegraphics[width=6.5in, height=2.5in]{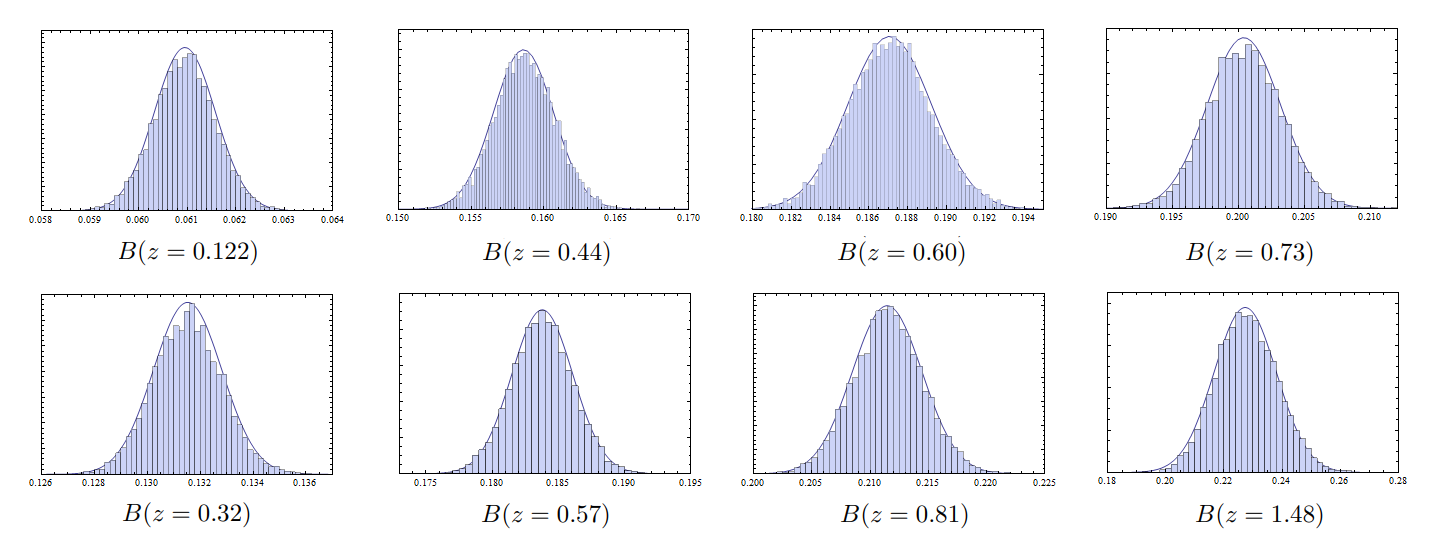}
\caption{Histograms of $B(z_i)$ \eqref{eq:Bfunc} at the redshifts $z_i$ of interest, see \eqref{eq:Bvector}. They have been obtained from the sampling of $10^4$ points with the procedure explained in Sec. \ref{sec:IOI1}. We also plot the Gaussian distributions built making use of \eqref{eq:Bvector}. One can see that the distributions are almost perfectly Gaussian.}\label{fig:histoB}
\end{center}
\end{figure*}

\section{Gaussian distribution for $B(z)$}\label{sec:appB}

The distribution for the values $B(z_i)$ \eqref{eq:Bfunc} at the redshifts $z_i$ of interest, needed to compute $\mu^{\rm SNIa}$ and $C^{\rm SNIa}$ (see Sec. \ref{sec:IOI1}), is Gaussian (cf. Fig. \ref{fig:histoB}). The corresponding mean and covariance matrix read,

\begin{widetext}
\begin{equation}\label{eq:Bvector}
\mu_B=\begin{pmatrix}
B(z=0.122)\\
B(z=0.44)\\
B(z=0.60)\\
B(z=0.73)\\
B(z=0.32)\\
B(z=0.57)\\
B(z=0.81)\\
B(z=1.48)
\end{pmatrix}=\begin{pmatrix}
0.06094\\
0.15869\\
0.18705\\
0.20048\\
0.13150\\
0.18374\\
0.21156\\  
0.22756
\end{pmatrix}\quad C_B=10^{-8}\begin{pmatrix}
39.45 & 0.26 & 3.91 & -6.15 & 2.36 & 1.60 & -9.17 & -23.26\\
-& 410.30 & 6.99 & 37.58 & 8.21 & 33.21 & 0.72 & -49.87\\
-&-& 460.39 & 70.55 & -10.73 & 364.85 & 72.72 & 113.22\\
-&-&-& 784.12 & -26.46 & 119.12 & 291.36 & 51.58\\
-&-&-&-& 168.67 & -15.79 & -33.62 & -83.11\\
-&-&-&-&-& 521.70 & 96.06 & -112.98\\
-&-&-&-&-&-& 906.54 & 716.72\\
-&-&-&-&-&-&-& 10869.4
\end{pmatrix}\,.
\end{equation}
\end{widetext}


\section{Gaussian Processes to reconstruct the Hubble function}\label{sec:appC}

Gaussian processes (GPs) can be thought of as Gaussian distributions over functions \citep{RasmussenWilliams}. They can be used to reconstruct the most probable underlying continuous function describing a set of Gaussian-distributed data and obtain its associated confidence regions without assuming any parametrization for the aforesaid function. One has to assume, though, a concrete kernel, which is in charge of controlling the correlations between the points of the reconstructed function. A GP is defined by two objects: its mean function, $\mu(z)$, and its two-point covariance function $\mathcal{C}(z,z^\prime)$,
\begin{equation}\label{eq:GP1}
\xi(z)\sim \mathcal{GP} \left(\mu(z),\mathcal{C}(z,z^\prime)\right)\,.
\end{equation}
Any realization $\xi(z)$ of a GP is a continuous curve, and it is of course possible to compute the probability of finding a realization inside any region $\xi(z)\pm \Delta\xi(z)$. The covariance function $\mathcal{C}$ is defined as follows: 

\begin{figure*}[t!]
\begin{center}
\includegraphics[width=6in, height=4in]{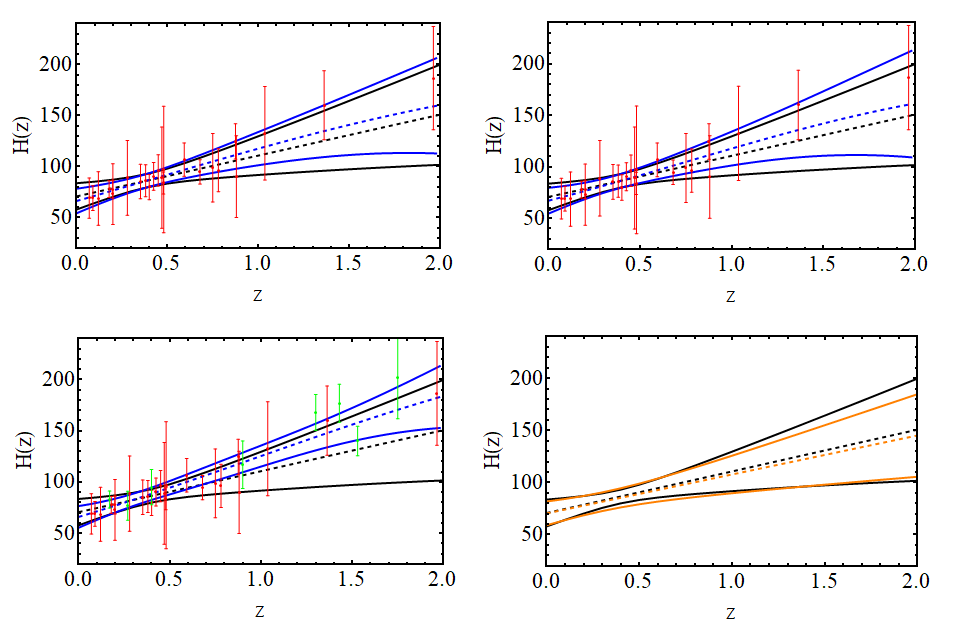}
\caption{Reconstructed shape of the Hubble function using Gaussian Processes under different kernels and CCH data sets. $H(z)$ is expressed in km/s/Mpc. The data points employed in our main analysis are shown in red, whereas those from \citep{Simon:2004tf} appearing in the lower left plot are in green. The dotted lines are the mean curves of the reconstructed Hubble function and the solid ones are the borders of the corresponding $2\sigma$ confidence regions. In all plots we show the results obtained with the Gaussian kernel \eqref{eq:Gaussiankernel} performing the full Bayesian analysis for the hyperparameters (in black). We compare it with the reconstructed Hubble function (in blue) obtained by minimizing the marginal likelihood \eqref{eq:logL} and using: the Gaussian kernel \eqref{eq:Gaussiankernel} (upper left plot); the Cauchy kernel \eqref{eq:Cauchykernel} (upper right plot); and the Gaussian kernel, considering also the data from \citep{Simon:2004tf} (lower left plot). The values of $H_0$ obtained for these three cases read $(66.16\pm 6.02,67.05\pm 6.26,66.02\pm 5.31)$ km/s/Mpc, respectively, at $1\sigma$ c.l., whereas for the main scenario (in black) we obtain $(70.72\pm 6.44)$ km/s/Mpc. The orange curve in the lower right plot is the reconstruction obtained as the black one, but using the covariance matrix of the CCH data (see Sec. \ref{sec:CCH}). In this case we obtain $H_0=(70.36\pm 5.69)$ km/s/Mpc.}\label{fig:Hrec}
\end{center}
\end{figure*}

\begin{widetext}
\begin{equation}
\mathcal{C}(z_1,z_2) =
\left\{
	\begin{array}{ll}
		\mathcal{K}(z_1,z_2)  & \mbox{if } z_1{\rm \,or}\,z_2\, \mbox{are not among the data points} \\
		\mathcal{D}(z_1,z_2)+\mathcal{K}(z_1,z_2) & \mbox{otherwise}
	\end{array}
\right.
\end{equation}
\end{widetext}
where $\mathcal{D}$ is the covariance matrix of the data points and $\mathcal{K}$ is the kernel function. Two of the most famous kernels are:

\begin{itemize}
\item The Gaussian kernel:
\begin{equation}\label{eq:Gaussiankernel}
\mathcal{K}(z_1,z_2)=\sigma_f^2e^{-\frac{1}{2}\left(\frac{z_1-z_2}{l_f}\right)^2}\,.
\end{equation}
\item The Cauchy kernel:
\begin{equation}\label{eq:Cauchykernel}
\mathcal{K}(z_1,z_2)=\frac{\sigma_f^2l_f}{(z_1-z_2)^2+l_f^2}\,.
\end{equation}
\end{itemize}
$\sigma_f$ and $l_f$ are the so-called hyperparameters of the kernel function. The first one controls the uncertainties' size, whereas the second regulates the scope of the correlations in $z$, i.e. for distances $|z_1-z_2|\gg l_f$ the values of the function at $z_1$ and $z_2$ are very uncorrelated. The reconstructed function depends on the values of the hyperparameters and, in principle, also on our choice of the kernel. We will study the impact of the latter below. For now, let us see how to select $\sigma_f$ and $l_f$ assuming a concrete kernel. In the GP philosophy, our data set is conceived just as a realization of the Gaussian process. The hyperparameters are usually chosen so as to maximize the probability of the GP to produce our data set. If we marginalize the GP \eqref{eq:GP1} over the points at $z^*$ (with no data) we get the following multivariate normal distribution, 
\begin{equation}\label{eq:GP2}
\vec{\xi}\sim \mathcal{N} \left(\{\mu_i(\tilde{z}_i)\},\mathcal{C}\right)\,,
\end{equation}
where $i=1,...,N$, with $N$ being the dimension of the vector of data points $\vec{y}\equiv\{\tilde{z}_i,y_i\}$ at our disposal. $\mu_i(\tilde{z}_i)$ can be set e.g. to $\vec{0}$ $\forall{i}$, since the result is almost insensitive to this. Thus, the hyperparameters will be obtained upon the minimization of 
\begin{equation}\label{eq:logL}
-2\ln\mathcal{L}(\sigma_f,l_f)=N\ln(2\pi)+\ln|\mathcal{C}(\sigma_f,l_f)|+\vec{y}^T\mathcal{C}^{-1}(\sigma_f,l_f)\vec{y}\,,
\end{equation}
with $\mathcal{L}$ being here the marginal likelihood and $|\mathcal{C}|$ the determinant of $\mathcal{C}$. Using \eqref{eq:GP1} one can compute now the conditional probability of finding a given realization of the Gaussian process in the case in which $\xi(\tilde{z}_i)=y_i(\tilde{z}_i)$. The resulting mean and covariance function extracted from the conditioned GP read, respectively,
\begin{equation}
\bar{\xi}(z^*)=\sum_{i,j=1}^{N}\mathcal{C}^{-1}(\tilde{z}_i,\tilde{z}_j)y(\tilde{z}_j)\mathcal{K}(\tilde{z}_i,z^*)\,,
\end{equation}
\begin{equation}
{\rm Cov}(z^*_1,z^*_2)=\mathcal{K}(z_1^*,z_2^*)-\sum_{i,j=1}^{N}\mathcal{C}^{-1}(\tilde{z}_i,\tilde{z}_j)\mathcal{K}(\tilde{z}_i,z_1^*)\mathcal{K}(\tilde{z}_j,z_2^*)\,.
\end{equation}
A complete Bayesian analysis would demand to marginalize over the hyperparameters, rather than optimizing Eq. \eqref{eq:logL}. We prefer to consider in our main analysis the full distribution for $(l_f,\sigma_f)$. In Fig. \ref{fig:Hrec} we show a plot comparing the reconstructed shape of $H(z)$ obtained with the Gaussian kernel \eqref{eq:Gaussiankernel} by both, optimizing and performing the complete Bayesian analysis. We also compare these results with those obtained with the optimization procedure using the Cauchy kernel \eqref{eq:Cauchykernel}, and using the optimized Gaussian kernel, but adding the data points from \citep{Simon:2004tf} that we have excluded in our main analysis. In addition, we also study the impact of the covariance matrix of the CCH data \cite{Moresco:2020fbm}.  Figure \ref{fig:Hrec} allows us to check that: (i) the impact of using the full Bayesian approach is not huge, but it is safer to propagate the uncertainties on the hyperparameters; (ii) our reconstructed Hubble function is stable under the change of the kernel; (iii) the effect that causes the removal of the data points from \cite{Simon:2004tf} is sizable, but the reconstructed shape is still compatible at $1\sigma$ c.l. with the one employed by us in the evaluation of $L(\vec{\theta})$ \eqref{eq:L}; (iv) the impact of the covariance matrix of the CCH data is small. For previous reconstructions of $H(z)$ with CCH data see e.g. \citep{Busti:2014dua,Verde:2014qea,Yu:2017iju,Haridasu:2018gqm,Gomez-Valent:2018hwc}.

We use our reconstructed Hubble function to obtain the data vector $\mu^{\rm CCH}$ and its corresponding covariance matrix $C^{\rm CCH}$, which are needed to compute $L_1$ \eqref{eq:L1} (see Sec. \ref{sec:IOI1} for details). In our main analyses we employ

\begin{widetext}
\begin{equation}
\mu^{\rm CCH}= \begin{pmatrix}
H(z=0.122)\\
H(z=0.44)\\
H(z=0.60)\\
H(z=0.73)\\
H(z=0.32)\\
H(z=0.57)\\
H(z=1.48)
\end{pmatrix}=
\begin{pmatrix}
75.55\\
88.20\\
94.60\\
99.81\\
83.41\\
93.40\\
129.94
\end{pmatrix}\,,\qquad C^{\rm CCH}= \begin{pmatrix}
24.89 & 7.57 & -1.12 & -8.15 & 14.10 & 0.51 & -47.92 \\
- & 11.33 & 13.11 & 14.50 & 9.95 & 12.78 & 21.69 \\\
- & - & 20.21 & 25.95 & 7.75 & 18.88 & 58.11 \\
- &- & -& 35.28 & 5.93 & 23.80 & 88.34 \\
- &- &- &- & 11.55& 8.17 & -5.03 \\
- &- &- &- &- & 17.74 & 51.22\\
- & - & -&- & -&- & 270.45\\
\end{pmatrix}\,,
\end{equation}
\end{widetext}
whereas in the case in which we use the covariance from \cite{Moresco:2020fbm} we consider

\begin{widetext}
\begin{equation}
\mu^{\rm CCH}= \begin{pmatrix}
H(z=0.122)\\
H(z=0.44)\\
H(z=0.60)\\
H(z=0.73)\\
H(z=0.32)\\
H(z=0.57)\\
H(z=1.48)
\end{pmatrix}=
\begin{pmatrix}
74.91\\
86.79\\
92.77\\
97.63\\
82.31\\
91.65\\
125.65
\end{pmatrix}\,,\qquad C^{\rm CCH}= \begin{pmatrix}
24.54 & 16.79 & 12.91 & 9.75 & 19.72 & 13.64 & -8.51 \\
- & 22.06 & 24.68 & 26.80 & 20.08 & 24.19 & 38.99 \\\
- & - & 30.63 & 35.41 & 20.25 & 29.52 & 63.10 \\
- &- & -& 42.42 & 20.37 & 33.80 & 82.76 \\
- &- &- &- & 19.97 & 20.21 & 21.01 \\
- &- &- &- &- & 28.52 & 58.57\\
- & - & -&- & -&- & 197.21\\
\end{pmatrix}\,.
\end{equation}
\end{widetext}
The data vectors and covariance matrices are expressed in km/s/Mpc and $({\rm km/s/Mpc})^2$, respectively.


\bibliographystyle{apsrev4-1}
\bibliography{Measurement_rd_M}

\begin{thebibliography}{97}%
\makeatletter
\providecommand \@ifxundefined [1]{%
 \@ifx{#1\undefined}
}%
\providecommand \@ifnum [1]{%
 \ifnum #1\expandafter \@firstoftwo
 \else \expandafter \@secondoftwo
 \fi
}%
\providecommand \@ifx [1]{%
 \ifx #1\expandafter \@firstoftwo
 \else \expandafter \@secondoftwo
 \fi
}%
\providecommand \natexlab [1]{#1}%
\providecommand \enquote  [1]{``#1''}%
\providecommand \bibnamefont  [1]{#1}%
\providecommand \bibfnamefont [1]{#1}%
\providecommand \citenamefont [1]{#1}%
\providecommand \href@noop [0]{\@secondoftwo}%
\providecommand \href [0]{\begingroup \@sanitize@url \@href}%
\providecommand \@href[1]{\@@startlink{#1}\@@href}%
\providecommand \@@href[1]{\endgroup#1\@@endlink}%
\providecommand \@sanitize@url [0]{\catcode `\\12\catcode `\$12\catcode
  `\&12\catcode `\#12\catcode `\^12\catcode `\_12\catcode `\%12\relax}%
\providecommand \@@startlink[1]{}%
\providecommand \@@endlink[0]{}%
\providecommand \url  [0]{\begingroup\@sanitize@url \@url }%
\providecommand \@url [1]{\endgroup\@href {#1}{\urlprefix }}%
\providecommand \urlprefix  [0]{URL }%
\providecommand \Eprint [0]{\href }%
\providecommand \doibase [0]{http://dx.doi.org/}%
\providecommand \selectlanguage [0]{\@gobble}%
\providecommand \bibinfo  [0]{\@secondoftwo}%
\providecommand \bibfield  [0]{\@secondoftwo}%
\providecommand \translation [1]{[#1]}%
\providecommand \BibitemOpen [0]{}%
\providecommand \bibitemStop [0]{}%
\providecommand \bibitemNoStop [0]{.\EOS\space}%
\providecommand \EOS [0]{\spacefactor3000\relax}%
\providecommand \BibitemShut  [1]{\csname bibitem#1\endcsname}%
\let\auto@bib@innerbib\@empty
\bibitem [{\citenamefont {Hubble}(1929)}]{Hubble:1929}%
  \BibitemOpen
  \bibfield  {author} {\bibinfo {author} {\bibfnamefont {E.}~\bibnamefont
  {Hubble}},\ }\href {\doibase 10.1073/pnas.15.3.168} {\bibfield  {journal}
  {\bibinfo  {journal} {Proc. Nat. Acad. Sci.}\ }\textbf {\bibinfo {volume}
  {15}},\ \bibinfo {pages} {168} (\bibinfo {year} {1929})}\BibitemShut
  {NoStop}%
\bibitem [{\citenamefont {Riess}\ \emph
  {et~al.}(2021{\natexlab{a}})\citenamefont {Riess} \emph
  {et~al.}}]{Riess:2021jrx}%
  \BibitemOpen
  \bibfield  {author} {\bibinfo {author} {\bibfnamefont {A.~G.}\ \bibnamefont
  {Riess}} \emph {et~al.},\ }\href@noop {} {\  (\bibinfo {year}
  {2021}{\natexlab{a}})},\ \Eprint {http://arxiv.org/abs/2112.04510}
  {arXiv:2112.04510 [astro-ph.CO]} \BibitemShut {NoStop}%
\bibitem [{\citenamefont {Aghanim}\ \emph {et~al.}(2020)\citenamefont {Aghanim}
  \emph {et~al.}}]{Planck:2018vyg}%
  \BibitemOpen
  \bibfield  {author} {\bibinfo {author} {\bibfnamefont {N.}~\bibnamefont
  {Aghanim}} \emph {et~al.} (\bibinfo {collaboration} {Planck}),\ }\href
  {\doibase 10.1051/0004-6361/201833910} {\bibfield  {journal} {\bibinfo
  {journal} {Astron. Astrophys.}\ }\textbf {\bibinfo {volume} {641}},\ \bibinfo
  {pages} {A6} (\bibinfo {year} {2020})},\ \Eprint
  {http://arxiv.org/abs/1807.06209} {arXiv:1807.06209 [astro-ph.CO]}
  \BibitemShut {NoStop}%
\bibitem [{\citenamefont {Camarena}\ and\ \citenamefont
  {Marra}(2020{\natexlab{a}})}]{Camarena:2019moy}%
  \BibitemOpen
  \bibfield  {author} {\bibinfo {author} {\bibfnamefont {D.}~\bibnamefont
  {Camarena}}\ and\ \bibinfo {author} {\bibfnamefont {V.}~\bibnamefont
  {Marra}},\ }\href {\doibase 10.1103/PhysRevResearch.2.013028} {\bibfield
  {journal} {\bibinfo  {journal} {Phys. Rev. Res.}\ }\textbf {\bibinfo {volume}
  {2}},\ \bibinfo {pages} {013028} (\bibinfo {year} {2020}{\natexlab{a}})},\
  \Eprint {http://arxiv.org/abs/1906.11814} {arXiv:1906.11814 [astro-ph.CO]}
  \BibitemShut {NoStop}%
\bibitem [{\citenamefont {Efstathiou}(2021)}]{Efstathiou:2021ocp}%
  \BibitemOpen
  \bibfield  {author} {\bibinfo {author} {\bibfnamefont {G.}~\bibnamefont
  {Efstathiou}},\ }\href {\doibase 10.1093/mnras/stab1588} {\bibfield
  {journal} {\bibinfo  {journal} {Mon. Not. Roy. Astron. Soc.}\ }\textbf
  {\bibinfo {volume} {505}},\ \bibinfo {pages} {3866} (\bibinfo {year}
  {2021})},\ \Eprint {http://arxiv.org/abs/2103.08723} {arXiv:2103.08723
  [astro-ph.CO]} \BibitemShut {NoStop}%
\bibitem [{\citenamefont {Freedman}\ \emph {et~al.}(2019)\citenamefont
  {Freedman} \emph {et~al.}}]{Freedman:2019jwv}%
  \BibitemOpen
  \bibfield  {author} {\bibinfo {author} {\bibfnamefont {W.~L.}\ \bibnamefont
  {Freedman}} \emph {et~al.},\ }\href {\doibase 10.3847/1538-4357/ab2f73} {\
  (\bibinfo {year} {2019}),\ 10.3847/1538-4357/ab2f73},\ \Eprint
  {http://arxiv.org/abs/1907.05922} {arXiv:1907.05922 [astro-ph.CO]}
  \BibitemShut {NoStop}%
\bibitem [{\citenamefont {Freedman}(2021)}]{Freedman:2021ahq}%
  \BibitemOpen
  \bibfield  {author} {\bibinfo {author} {\bibfnamefont {W.~L.}\ \bibnamefont
  {Freedman}},\ }\href {\doibase 10.3847/1538-4357/ac0e95} {\bibfield
  {journal} {\bibinfo  {journal} {Astrophys. J.}\ }\textbf {\bibinfo {volume}
  {919}},\ \bibinfo {pages} {16} (\bibinfo {year} {2021})},\ \Eprint
  {http://arxiv.org/abs/2106.15656} {arXiv:2106.15656 [astro-ph.CO]}
  \BibitemShut {NoStop}%
\bibitem [{\citenamefont {Efstathiou}(2020)}]{Efstathiou:2020wxn}%
  \BibitemOpen
  \bibfield  {author} {\bibinfo {author} {\bibfnamefont {G.}~\bibnamefont
  {Efstathiou}},\ }\href@noop {} {\  (\bibinfo {year} {2020})},\ \Eprint
  {http://arxiv.org/abs/2007.10716} {arXiv:2007.10716 [astro-ph.CO]}
  \BibitemShut {NoStop}%
\bibitem [{\citenamefont {Mortsell}\ \emph {et~al.}(2021)\citenamefont
  {Mortsell}, \citenamefont {Goobar}, \citenamefont {Johansson},\ and\
  \citenamefont {Dhawan}}]{Mortsell:2021nzg}%
  \BibitemOpen
  \bibfield  {author} {\bibinfo {author} {\bibfnamefont {E.}~\bibnamefont
  {Mortsell}}, \bibinfo {author} {\bibfnamefont {A.}~\bibnamefont {Goobar}},
  \bibinfo {author} {\bibfnamefont {J.}~\bibnamefont {Johansson}}, \ and\
  \bibinfo {author} {\bibfnamefont {S.}~\bibnamefont {Dhawan}},\ }\href@noop {}
  {\  (\bibinfo {year} {2021})},\ \Eprint {http://arxiv.org/abs/2105.11461}
  {arXiv:2105.11461 [astro-ph.CO]} \BibitemShut {NoStop}%
\bibitem [{\citenamefont {Riess}\ \emph {et~al.}(2009)\citenamefont {Riess}
  \emph {et~al.}}]{Riess:2009pu}%
  \BibitemOpen
  \bibfield  {author} {\bibinfo {author} {\bibfnamefont {A.~G.}\ \bibnamefont
  {Riess}} \emph {et~al.},\ }\href {\doibase 10.1088/0004-637X/699/1/539}
  {\bibfield  {journal} {\bibinfo  {journal} {Astrophys. J.}\ }\textbf
  {\bibinfo {volume} {699}},\ \bibinfo {pages} {539} (\bibinfo {year}
  {2009})},\ \Eprint {http://arxiv.org/abs/0905.0695} {arXiv:0905.0695
  [astro-ph.CO]} \BibitemShut {NoStop}%
\bibitem [{\citenamefont {Riess}\ \emph {et~al.}(2016)\citenamefont {Riess}
  \emph {et~al.}}]{Riess:2016jrr}%
  \BibitemOpen
  \bibfield  {author} {\bibinfo {author} {\bibfnamefont {A.~G.}\ \bibnamefont
  {Riess}} \emph {et~al.},\ }\href {\doibase 10.3847/0004-637X/826/1/56}
  {\bibfield  {journal} {\bibinfo  {journal} {Astrophys. J.}\ }\textbf
  {\bibinfo {volume} {826}},\ \bibinfo {pages} {56} (\bibinfo {year} {2016})},\
  \Eprint {http://arxiv.org/abs/1604.01424} {arXiv:1604.01424 [astro-ph.CO]}
  \BibitemShut {NoStop}%
\bibitem [{\citenamefont {Riess}\ \emph
  {et~al.}(2021{\natexlab{b}})\citenamefont {Riess}, \citenamefont {Casertano},
  \citenamefont {Yuan}, \citenamefont {Bowers}, \citenamefont {Macri},
  \citenamefont {Zinn},\ and\ \citenamefont {Scolnic}}]{Riess:2020fzl}%
  \BibitemOpen
  \bibfield  {author} {\bibinfo {author} {\bibfnamefont {A.~G.}\ \bibnamefont
  {Riess}}, \bibinfo {author} {\bibfnamefont {S.}~\bibnamefont {Casertano}},
  \bibinfo {author} {\bibfnamefont {W.}~\bibnamefont {Yuan}}, \bibinfo {author}
  {\bibfnamefont {J.~B.}\ \bibnamefont {Bowers}}, \bibinfo {author}
  {\bibfnamefont {L.}~\bibnamefont {Macri}}, \bibinfo {author} {\bibfnamefont
  {J.~C.}\ \bibnamefont {Zinn}}, \ and\ \bibinfo {author} {\bibfnamefont
  {D.}~\bibnamefont {Scolnic}},\ }\href {\doibase 10.3847/2041-8213/abdbaf}
  {\bibfield  {journal} {\bibinfo  {journal} {Astrophys. J. Lett.}\ }\textbf
  {\bibinfo {volume} {908}},\ \bibinfo {pages} {L6} (\bibinfo {year}
  {2021}{\natexlab{b}})},\ \Eprint {http://arxiv.org/abs/2012.08534}
  {arXiv:2012.08534 [astro-ph.CO]} \BibitemShut {NoStop}%
\bibitem [{\citenamefont {Hinshaw}\ \emph {et~al.}(2013)\citenamefont {Hinshaw}
  \emph {et~al.}}]{WMAP:2012nax}%
  \BibitemOpen
  \bibfield  {author} {\bibinfo {author} {\bibfnamefont {G.}~\bibnamefont
  {Hinshaw}} \emph {et~al.} (\bibinfo {collaboration} {WMAP}),\ }\href
  {\doibase 10.1088/0067-0049/208/2/19} {\bibfield  {journal} {\bibinfo
  {journal} {Astrophys. J. Suppl.}\ }\textbf {\bibinfo {volume} {208}},\
  \bibinfo {pages} {19} (\bibinfo {year} {2013})},\ \Eprint
  {http://arxiv.org/abs/1212.5226} {arXiv:1212.5226 [astro-ph.CO]} \BibitemShut
  {NoStop}%
\bibitem [{\citenamefont {Ade}\ \emph {et~al.}(2016)\citenamefont {Ade} \emph
  {et~al.}}]{Planck:2015fie}%
  \BibitemOpen
  \bibfield  {author} {\bibinfo {author} {\bibfnamefont {P.~A.~R.}\
  \bibnamefont {Ade}} \emph {et~al.} (\bibinfo {collaboration} {Planck}),\
  }\href {\doibase 10.1051/0004-6361/201525830} {\bibfield  {journal} {\bibinfo
   {journal} {Astron. Astrophys.}\ }\textbf {\bibinfo {volume} {594}},\
  \bibinfo {pages} {A13} (\bibinfo {year} {2016})},\ \Eprint
  {http://arxiv.org/abs/1502.01589} {arXiv:1502.01589 [astro-ph.CO]}
  \BibitemShut {NoStop}%
\bibitem [{\citenamefont {Aylor}\ \emph {et~al.}(2017)\citenamefont {Aylor}
  \emph {et~al.}}]{SPT:2017sjt}%
  \BibitemOpen
  \bibfield  {author} {\bibinfo {author} {\bibfnamefont {K.}~\bibnamefont
  {Aylor}} \emph {et~al.} (\bibinfo {collaboration} {SPT}),\ }\href {\doibase
  10.3847/1538-4357/aa947b} {\bibfield  {journal} {\bibinfo  {journal}
  {Astrophys. J.}\ }\textbf {\bibinfo {volume} {850}},\ \bibinfo {pages} {101}
  (\bibinfo {year} {2017})},\ \Eprint {http://arxiv.org/abs/1706.10286}
  {arXiv:1706.10286 [astro-ph.CO]} \BibitemShut {NoStop}%
\bibitem [{\citenamefont {Aiola}\ \emph {et~al.}(2020)\citenamefont {Aiola}
  \emph {et~al.}}]{ACT:2020gnv}%
  \BibitemOpen
  \bibfield  {author} {\bibinfo {author} {\bibfnamefont {S.}~\bibnamefont
  {Aiola}} \emph {et~al.} (\bibinfo {collaboration} {ACT}),\ }\href {\doibase
  10.1088/1475-7516/2020/12/047} {\bibfield  {journal} {\bibinfo  {journal}
  {JCAP}\ }\textbf {\bibinfo {volume} {12}},\ \bibinfo {pages} {047} (\bibinfo
  {year} {2020})},\ \Eprint {http://arxiv.org/abs/2007.07288} {arXiv:2007.07288
  [astro-ph.CO]} \BibitemShut {NoStop}%
\bibitem [{\citenamefont {Addison}\ \emph {et~al.}(2018)\citenamefont
  {Addison}, \citenamefont {Watts}, \citenamefont {Bennett}, \citenamefont
  {Halpern}, \citenamefont {Hinshaw},\ and\ \citenamefont
  {Weiland}}]{Addison:2017fdm}%
  \BibitemOpen
  \bibfield  {author} {\bibinfo {author} {\bibfnamefont {G.~E.}\ \bibnamefont
  {Addison}}, \bibinfo {author} {\bibfnamefont {D.~J.}\ \bibnamefont {Watts}},
  \bibinfo {author} {\bibfnamefont {C.~L.}\ \bibnamefont {Bennett}}, \bibinfo
  {author} {\bibfnamefont {M.}~\bibnamefont {Halpern}}, \bibinfo {author}
  {\bibfnamefont {G.}~\bibnamefont {Hinshaw}}, \ and\ \bibinfo {author}
  {\bibfnamefont {J.~L.}\ \bibnamefont {Weiland}},\ }\href {\doibase
  10.3847/1538-4357/aaa1ed} {\bibfield  {journal} {\bibinfo  {journal}
  {Astrophys. J.}\ }\textbf {\bibinfo {volume} {853}},\ \bibinfo {pages} {119}
  (\bibinfo {year} {2018})},\ \Eprint {http://arxiv.org/abs/1707.06547}
  {arXiv:1707.06547 [astro-ph.CO]} \BibitemShut {NoStop}%
\bibitem [{\citenamefont {Cuceu}\ \emph {et~al.}(2019)\citenamefont {Cuceu},
  \citenamefont {Farr}, \citenamefont {Lemos},\ and\ \citenamefont
  {Font-Ribera}}]{Cuceu:2019for}%
  \BibitemOpen
  \bibfield  {author} {\bibinfo {author} {\bibfnamefont {A.}~\bibnamefont
  {Cuceu}}, \bibinfo {author} {\bibfnamefont {J.}~\bibnamefont {Farr}},
  \bibinfo {author} {\bibfnamefont {P.}~\bibnamefont {Lemos}}, \ and\ \bibinfo
  {author} {\bibfnamefont {A.}~\bibnamefont {Font-Ribera}},\ }\href {\doibase
  10.1088/1475-7516/2019/10/044} {\bibfield  {journal} {\bibinfo  {journal}
  {JCAP}\ }\textbf {\bibinfo {volume} {10}},\ \bibinfo {pages} {044} (\bibinfo
  {year} {2019})},\ \Eprint {http://arxiv.org/abs/1906.11628} {arXiv:1906.11628
  [astro-ph.CO]} \BibitemShut {NoStop}%
\bibitem [{\citenamefont {Aubourg}\ \emph {et~al.}(2015)\citenamefont {Aubourg}
  \emph {et~al.}}]{Aubourg:2014yra}%
  \BibitemOpen
  \bibfield  {author} {\bibinfo {author} {\bibfnamefont {E.}~\bibnamefont
  {Aubourg}} \emph {et~al.},\ }\href {\doibase 10.1103/PhysRevD.92.123516}
  {\bibfield  {journal} {\bibinfo  {journal} {Phys. Rev. D}\ }\textbf {\bibinfo
  {volume} {92}},\ \bibinfo {pages} {123516} (\bibinfo {year} {2015})},\
  \Eprint {http://arxiv.org/abs/1411.1074} {arXiv:1411.1074 [astro-ph.CO]}
  \BibitemShut {NoStop}%
\bibitem [{\citenamefont {Cuesta}\ \emph {et~al.}(2015)\citenamefont {Cuesta},
  \citenamefont {Verde}, \citenamefont {Riess},\ and\ \citenamefont
  {Jim\'enez}}]{Cuesta:2014asa}%
  \BibitemOpen
  \bibfield  {author} {\bibinfo {author} {\bibfnamefont {A.~J.}\ \bibnamefont
  {Cuesta}}, \bibinfo {author} {\bibfnamefont {L.}~\bibnamefont {Verde}},
  \bibinfo {author} {\bibfnamefont {A.}~\bibnamefont {Riess}}, \ and\ \bibinfo
  {author} {\bibfnamefont {R.}~\bibnamefont {Jim\'enez}},\ }\href {\doibase
  10.1093/mnras/stv261} {\bibfield  {journal} {\bibinfo  {journal} {Mon. Not.
  Roy. Astron. Soc.}\ }\textbf {\bibinfo {volume} {448}},\ \bibinfo {pages}
  {3463} (\bibinfo {year} {2015})},\ \Eprint {http://arxiv.org/abs/1411.1094}
  {arXiv:1411.1094 [astro-ph.CO]} \BibitemShut {NoStop}%
\bibitem [{\citenamefont {Feeney}\ \emph {et~al.}(2019)\citenamefont {Feeney},
  \citenamefont {Peiris}, \citenamefont {Williamson}, \citenamefont {Nissanke},
  \citenamefont {Mortlock}, \citenamefont {Alsing},\ and\ \citenamefont
  {Scolnic}}]{Feeney:2018mkj}%
  \BibitemOpen
  \bibfield  {author} {\bibinfo {author} {\bibfnamefont {S.~M.}\ \bibnamefont
  {Feeney}}, \bibinfo {author} {\bibfnamefont {H.~V.}\ \bibnamefont {Peiris}},
  \bibinfo {author} {\bibfnamefont {A.~R.}\ \bibnamefont {Williamson}},
  \bibinfo {author} {\bibfnamefont {S.~M.}\ \bibnamefont {Nissanke}}, \bibinfo
  {author} {\bibfnamefont {D.~J.}\ \bibnamefont {Mortlock}}, \bibinfo {author}
  {\bibfnamefont {J.}~\bibnamefont {Alsing}}, \ and\ \bibinfo {author}
  {\bibfnamefont {D.}~\bibnamefont {Scolnic}},\ }\href {\doibase
  10.1103/PhysRevLett.122.061105} {\bibfield  {journal} {\bibinfo  {journal}
  {Phys. Rev. Lett.}\ }\textbf {\bibinfo {volume} {122}},\ \bibinfo {pages}
  {061105} (\bibinfo {year} {2019})},\ \Eprint
  {http://arxiv.org/abs/1802.03404} {arXiv:1802.03404 [astro-ph.CO]}
  \BibitemShut {NoStop}%
\bibitem [{\citenamefont {Camarena}\ and\ \citenamefont
  {Marra}(2020{\natexlab{b}})}]{Camarena:2019rmj}%
  \BibitemOpen
  \bibfield  {author} {\bibinfo {author} {\bibfnamefont {D.}~\bibnamefont
  {Camarena}}\ and\ \bibinfo {author} {\bibfnamefont {V.}~\bibnamefont
  {Marra}},\ }\href {\doibase 10.1093/mnras/staa770} {\bibfield  {journal}
  {\bibinfo  {journal} {Mon. Not. Roy. Astron. Soc.}\ }\textbf {\bibinfo
  {volume} {495}},\ \bibinfo {pages} {2630} (\bibinfo {year}
  {2020}{\natexlab{b}})},\ \Eprint {http://arxiv.org/abs/1910.14125}
  {arXiv:1910.14125 [astro-ph.CO]} \BibitemShut {NoStop}%
\bibitem [{\citenamefont {Abbott}\ \emph {et~al.}(2017)\citenamefont {Abbott}
  \emph {et~al.}}]{LIGOScientific:2017adf}%
  \BibitemOpen
  \bibfield  {author} {\bibinfo {author} {\bibfnamefont {B.~P.}\ \bibnamefont
  {Abbott}} \emph {et~al.} (\bibinfo {collaboration} {LIGO Scientific, Virgo,
  1M2H, Dark Energy Camera GW-E, DES, DLT40, Las Cumbres Observatory, VINROUGE,
  MASTER}),\ }\href {\doibase 10.1038/nature24471} {\bibfield  {journal}
  {\bibinfo  {journal} {Nature}\ }\textbf {\bibinfo {volume} {551}},\ \bibinfo
  {pages} {85} (\bibinfo {year} {2017})},\ \Eprint
  {http://arxiv.org/abs/1710.05835} {arXiv:1710.05835 [astro-ph.CO]}
  \BibitemShut {NoStop}%
\bibitem [{\citenamefont {Palmese}\ \emph {et~al.}(2021)\citenamefont
  {Palmese}, \citenamefont {Bom}, \citenamefont {Mucesh},\ and\ \citenamefont
  {Hartley}}]{Palmese:2021mjm}%
  \BibitemOpen
  \bibfield  {author} {\bibinfo {author} {\bibfnamefont {A.}~\bibnamefont
  {Palmese}}, \bibinfo {author} {\bibfnamefont {C.~R.}\ \bibnamefont {Bom}},
  \bibinfo {author} {\bibfnamefont {S.}~\bibnamefont {Mucesh}}, \ and\ \bibinfo
  {author} {\bibfnamefont {W.~G.}\ \bibnamefont {Hartley}},\ }\href@noop {} {\
  (\bibinfo {year} {2021})},\ \Eprint {http://arxiv.org/abs/2111.06445}
  {arXiv:2111.06445 [astro-ph.CO]} \BibitemShut {NoStop}%
\bibitem [{\citenamefont {Denzel}\ \emph {et~al.}(2021)\citenamefont {Denzel},
  \citenamefont {Coles}, \citenamefont {Saha},\ and\ \citenamefont
  {Williams}}]{Denzel:2020zuq}%
  \BibitemOpen
  \bibfield  {author} {\bibinfo {author} {\bibfnamefont {P.}~\bibnamefont
  {Denzel}}, \bibinfo {author} {\bibfnamefont {J.~P.}\ \bibnamefont {Coles}},
  \bibinfo {author} {\bibfnamefont {P.}~\bibnamefont {Saha}}, \ and\ \bibinfo
  {author} {\bibfnamefont {L.~L.~R.}\ \bibnamefont {Williams}},\ }\href
  {\doibase 10.1093/mnras/staa3603} {\bibfield  {journal} {\bibinfo  {journal}
  {Mon. Not. Roy. Astron. Soc.}\ }\textbf {\bibinfo {volume} {501}},\ \bibinfo
  {pages} {784} (\bibinfo {year} {2021})},\ \Eprint
  {http://arxiv.org/abs/2007.14398} {arXiv:2007.14398 [astro-ph.CO]}
  \BibitemShut {NoStop}%
\bibitem [{\citenamefont {Poulin}\ \emph {et~al.}(2019)\citenamefont {Poulin},
  \citenamefont {Smith}, \citenamefont {Karwal},\ and\ \citenamefont
  {Kamionkowski}}]{Poulin:2018cxd}%
  \BibitemOpen
  \bibfield  {author} {\bibinfo {author} {\bibfnamefont {V.}~\bibnamefont
  {Poulin}}, \bibinfo {author} {\bibfnamefont {T.~L.}\ \bibnamefont {Smith}},
  \bibinfo {author} {\bibfnamefont {T.}~\bibnamefont {Karwal}}, \ and\ \bibinfo
  {author} {\bibfnamefont {M.}~\bibnamefont {Kamionkowski}},\ }\href {\doibase
  10.1103/PhysRevLett.122.221301} {\bibfield  {journal} {\bibinfo  {journal}
  {Phys. Rev. Lett.}\ }\textbf {\bibinfo {volume} {122}},\ \bibinfo {pages}
  {221301} (\bibinfo {year} {2019})},\ \Eprint
  {http://arxiv.org/abs/1811.04083} {arXiv:1811.04083 [astro-ph.CO]}
  \BibitemShut {NoStop}%
\bibitem [{\citenamefont {Niedermann}\ and\ \citenamefont
  {Sloth}(2021)}]{Niedermann:2019olb}%
  \BibitemOpen
  \bibfield  {author} {\bibinfo {author} {\bibfnamefont {F.}~\bibnamefont
  {Niedermann}}\ and\ \bibinfo {author} {\bibfnamefont {M.~S.}\ \bibnamefont
  {Sloth}},\ }\href {\doibase 10.1103/PhysRevD.103.L041303} {\bibfield
  {journal} {\bibinfo  {journal} {Phys. Rev. D}\ }\textbf {\bibinfo {volume}
  {103}},\ \bibinfo {pages} {L041303} (\bibinfo {year} {2021})},\ \Eprint
  {http://arxiv.org/abs/1910.10739} {arXiv:1910.10739 [astro-ph.CO]}
  \BibitemShut {NoStop}%
\bibitem [{\citenamefont {Agrawal}\ \emph {et~al.}(2019)\citenamefont
  {Agrawal}, \citenamefont {Cyr-Racine}, \citenamefont {Pinner},\ and\
  \citenamefont {Randall}}]{Agrawal:2019lmo}%
  \BibitemOpen
  \bibfield  {author} {\bibinfo {author} {\bibfnamefont {P.}~\bibnamefont
  {Agrawal}}, \bibinfo {author} {\bibfnamefont {F.-Y.}\ \bibnamefont
  {Cyr-Racine}}, \bibinfo {author} {\bibfnamefont {D.}~\bibnamefont {Pinner}},
  \ and\ \bibinfo {author} {\bibfnamefont {L.}~\bibnamefont {Randall}},\
  }\href@noop {} {\  (\bibinfo {year} {2019})},\ \Eprint
  {http://arxiv.org/abs/1904.01016} {arXiv:1904.01016 [astro-ph.CO]}
  \BibitemShut {NoStop}%
\bibitem [{\citenamefont {G\'omez-Valent}\ \emph {et~al.}(2021)\citenamefont
  {G\'omez-Valent}, \citenamefont {Zheng}, \citenamefont {Amendola},
  \citenamefont {Pettorino},\ and\ \citenamefont
  {Wetterich}}]{Gomez-Valent:2021cbe}%
  \BibitemOpen
  \bibfield  {author} {\bibinfo {author} {\bibfnamefont {A.}~\bibnamefont
  {G\'omez-Valent}}, \bibinfo {author} {\bibfnamefont {Z.}~\bibnamefont
  {Zheng}}, \bibinfo {author} {\bibfnamefont {L.}~\bibnamefont {Amendola}},
  \bibinfo {author} {\bibfnamefont {V.}~\bibnamefont {Pettorino}}, \ and\
  \bibinfo {author} {\bibfnamefont {C.}~\bibnamefont {Wetterich}},\ }\href
  {\doibase 10.1103/PhysRevD.104.083536} {\bibfield  {journal} {\bibinfo
  {journal} {Phys. Rev. D}\ }\textbf {\bibinfo {volume} {104}},\ \bibinfo
  {pages} {083536} (\bibinfo {year} {2021})},\ \Eprint
  {http://arxiv.org/abs/2107.11065} {arXiv:2107.11065 [astro-ph.CO]}
  \BibitemShut {NoStop}%
\bibitem [{\citenamefont {Sol\`a~Peracaula}\ \emph {et~al.}(2019)\citenamefont
  {Sol\`a~Peracaula}, \citenamefont {G\'omez-Valent}, \citenamefont
  {de~Cruz~P\'erez},\ and\ \citenamefont
  {Moreno-Pulido}}]{SolaPeracaula:2019zsl}%
  \BibitemOpen
  \bibfield  {author} {\bibinfo {author} {\bibfnamefont {J.}~\bibnamefont
  {Sol\`a~Peracaula}}, \bibinfo {author} {\bibfnamefont {A.}~\bibnamefont
  {G\'omez-Valent}}, \bibinfo {author} {\bibfnamefont {J.}~\bibnamefont
  {de~Cruz~P\'erez}}, \ and\ \bibinfo {author} {\bibfnamefont {C.}~\bibnamefont
  {Moreno-Pulido}},\ }\href {\doibase 10.3847/2041-8213/ab53e9} {\bibfield
  {journal} {\bibinfo  {journal} {Astrophys. J. Lett.}\ }\textbf {\bibinfo
  {volume} {886}},\ \bibinfo {pages} {L6} (\bibinfo {year} {2019})},\ \Eprint
  {http://arxiv.org/abs/1909.02554} {arXiv:1909.02554 [astro-ph.CO]}
  \BibitemShut {NoStop}%
\bibitem [{\citenamefont {Sol\`a~Peracaula}\ \emph {et~al.}(2020)\citenamefont
  {Sol\`a~Peracaula}, \citenamefont {G\'omez-Valent}, \citenamefont
  {de~Cruz~P\'erez},\ and\ \citenamefont
  {Moreno-Pulido}}]{SolaPeracaula:2020vpg}%
  \BibitemOpen
  \bibfield  {author} {\bibinfo {author} {\bibfnamefont {J.}~\bibnamefont
  {Sol\`a~Peracaula}}, \bibinfo {author} {\bibfnamefont {A.}~\bibnamefont
  {G\'omez-Valent}}, \bibinfo {author} {\bibfnamefont {J.}~\bibnamefont
  {de~Cruz~P\'erez}}, \ and\ \bibinfo {author} {\bibfnamefont {C.}~\bibnamefont
  {Moreno-Pulido}},\ }\href {\doibase 10.1088/1361-6382/abbc43} {\bibfield
  {journal} {\bibinfo  {journal} {Class. Quant. Grav.}\ }\textbf {\bibinfo
  {volume} {37}},\ \bibinfo {pages} {245003} (\bibinfo {year} {2020})},\
  \Eprint {http://arxiv.org/abs/2006.04273} {arXiv:2006.04273 [astro-ph.CO]}
  \BibitemShut {NoStop}%
\bibitem [{\citenamefont {Braglia}\ \emph {et~al.}(2020)\citenamefont
  {Braglia}, \citenamefont {Ballardini}, \citenamefont {Emond}, \citenamefont
  {Finelli}, \citenamefont {Gumrukcuoglu}, \citenamefont {Koyama},\ and\
  \citenamefont {Paoletti}}]{Braglia:2020iik}%
  \BibitemOpen
  \bibfield  {author} {\bibinfo {author} {\bibfnamefont {M.}~\bibnamefont
  {Braglia}}, \bibinfo {author} {\bibfnamefont {M.}~\bibnamefont {Ballardini}},
  \bibinfo {author} {\bibfnamefont {W.~T.}\ \bibnamefont {Emond}}, \bibinfo
  {author} {\bibfnamefont {F.}~\bibnamefont {Finelli}}, \bibinfo {author}
  {\bibfnamefont {A.~E.}\ \bibnamefont {Gumrukcuoglu}}, \bibinfo {author}
  {\bibfnamefont {K.}~\bibnamefont {Koyama}}, \ and\ \bibinfo {author}
  {\bibfnamefont {D.}~\bibnamefont {Paoletti}},\ }\href {\doibase
  10.1103/PhysRevD.102.023529} {\bibfield  {journal} {\bibinfo  {journal}
  {Phys. Rev. D}\ }\textbf {\bibinfo {volume} {102}},\ \bibinfo {pages}
  {023529} (\bibinfo {year} {2020})},\ \Eprint
  {http://arxiv.org/abs/2004.11161} {arXiv:2004.11161 [astro-ph.CO]}
  \BibitemShut {NoStop}%
\bibitem [{\citenamefont {Braglia}\ \emph {et~al.}(2021)\citenamefont
  {Braglia}, \citenamefont {Ballardini}, \citenamefont {Finelli},\ and\
  \citenamefont {Koyama}}]{Braglia:2020auw}%
  \BibitemOpen
  \bibfield  {author} {\bibinfo {author} {\bibfnamefont {M.}~\bibnamefont
  {Braglia}}, \bibinfo {author} {\bibfnamefont {M.}~\bibnamefont {Ballardini}},
  \bibinfo {author} {\bibfnamefont {F.}~\bibnamefont {Finelli}}, \ and\
  \bibinfo {author} {\bibfnamefont {K.}~\bibnamefont {Koyama}},\ }\href
  {\doibase 10.1103/PhysRevD.103.043528} {\bibfield  {journal} {\bibinfo
  {journal} {Phys. Rev. D}\ }\textbf {\bibinfo {volume} {103}},\ \bibinfo
  {pages} {043528} (\bibinfo {year} {2021})},\ \Eprint
  {http://arxiv.org/abs/2011.12934} {arXiv:2011.12934 [astro-ph.CO]}
  \BibitemShut {NoStop}%
\bibitem [{\citenamefont {Jedamzik}\ and\ \citenamefont
  {Pogosian}(2020)}]{Jedamzik:2020krr}%
  \BibitemOpen
  \bibfield  {author} {\bibinfo {author} {\bibfnamefont {K.}~\bibnamefont
  {Jedamzik}}\ and\ \bibinfo {author} {\bibfnamefont {L.}~\bibnamefont
  {Pogosian}},\ }\href {\doibase 10.1103/PhysRevLett.125.181302} {\bibfield
  {journal} {\bibinfo  {journal} {Phys. Rev. Lett.}\ }\textbf {\bibinfo
  {volume} {125}},\ \bibinfo {pages} {181302} (\bibinfo {year} {2020})},\
  \Eprint {http://arxiv.org/abs/2004.09487} {arXiv:2004.09487 [astro-ph.CO]}
  \BibitemShut {NoStop}%
\bibitem [{\citenamefont {Liu}\ \emph {et~al.}(2020)\citenamefont {Liu},
  \citenamefont {Huang}, \citenamefont {Luo}, \citenamefont {Miao},
  \citenamefont {Singh},\ and\ \citenamefont {Huang}}]{Liu:2019awo}%
  \BibitemOpen
  \bibfield  {author} {\bibinfo {author} {\bibfnamefont {M.}~\bibnamefont
  {Liu}}, \bibinfo {author} {\bibfnamefont {Z.}~\bibnamefont {Huang}}, \bibinfo
  {author} {\bibfnamefont {X.}~\bibnamefont {Luo}}, \bibinfo {author}
  {\bibfnamefont {H.}~\bibnamefont {Miao}}, \bibinfo {author} {\bibfnamefont
  {N.~K.}\ \bibnamefont {Singh}}, \ and\ \bibinfo {author} {\bibfnamefont
  {L.}~\bibnamefont {Huang}},\ }\href {\doibase 10.1007/s11433-019-1509-5}
  {\bibfield  {journal} {\bibinfo  {journal} {Sci. China Phys. Mech. Astron.}\
  }\textbf {\bibinfo {volume} {63}},\ \bibinfo {pages} {290405} (\bibinfo
  {year} {2020})},\ \Eprint {http://arxiv.org/abs/1912.00190} {arXiv:1912.00190
  [astro-ph.CO]} \BibitemShut {NoStop}%
\bibitem [{\citenamefont {Sekiguchi}\ and\ \citenamefont
  {Takahashi}(2021)}]{Sekiguchi:2020teg}%
  \BibitemOpen
  \bibfield  {author} {\bibinfo {author} {\bibfnamefont {T.}~\bibnamefont
  {Sekiguchi}}\ and\ \bibinfo {author} {\bibfnamefont {T.}~\bibnamefont
  {Takahashi}},\ }\href {\doibase 10.1103/PhysRevD.103.083507} {\bibfield
  {journal} {\bibinfo  {journal} {Phys. Rev. D}\ }\textbf {\bibinfo {volume}
  {103}},\ \bibinfo {pages} {083507} (\bibinfo {year} {2021})},\ \Eprint
  {http://arxiv.org/abs/2007.03381} {arXiv:2007.03381 [astro-ph.CO]}
  \BibitemShut {NoStop}%
\bibitem [{\citenamefont {Sol\`a~Peracaula}\ \emph {et~al.}(2021)\citenamefont
  {Sol\`a~Peracaula}, \citenamefont {G\'omez-Valent}, \citenamefont
  {de~Cruz~P\'erez},\ and\ \citenamefont
  {Moreno-Pulido}}]{SolaPeracaula:2021gxi}%
  \BibitemOpen
  \bibfield  {author} {\bibinfo {author} {\bibfnamefont {J.}~\bibnamefont
  {Sol\`a~Peracaula}}, \bibinfo {author} {\bibfnamefont {A.}~\bibnamefont
  {G\'omez-Valent}}, \bibinfo {author} {\bibfnamefont {J.}~\bibnamefont
  {de~Cruz~P\'erez}}, \ and\ \bibinfo {author} {\bibfnamefont {C.}~\bibnamefont
  {Moreno-Pulido}},\ }\href {\doibase 10.1209/0295-5075/134/19001} {\bibfield
  {journal} {\bibinfo  {journal} {EPL}\ }\textbf {\bibinfo {volume} {134}},\
  \bibinfo {pages} {19001} (\bibinfo {year} {2021})},\ \Eprint
  {http://arxiv.org/abs/2102.12758} {arXiv:2102.12758 [astro-ph.CO]}
  \BibitemShut {NoStop}%
\bibitem [{\citenamefont {Verde}\ \emph
  {et~al.}(2017{\natexlab{a}})\citenamefont {Verde}, \citenamefont {Bellini},
  \citenamefont {Pigozzo}, \citenamefont {Heavens},\ and\ \citenamefont
  {Jim\'enez}}]{Verde:2016wmz}%
  \BibitemOpen
  \bibfield  {author} {\bibinfo {author} {\bibfnamefont {L.}~\bibnamefont
  {Verde}}, \bibinfo {author} {\bibfnamefont {E.}~\bibnamefont {Bellini}},
  \bibinfo {author} {\bibfnamefont {C.}~\bibnamefont {Pigozzo}}, \bibinfo
  {author} {\bibfnamefont {A.~F.}\ \bibnamefont {Heavens}}, \ and\ \bibinfo
  {author} {\bibfnamefont {R.}~\bibnamefont {Jim\'enez}},\ }\href {\doibase
  10.1088/1475-7516/2017/04/023} {\bibfield  {journal} {\bibinfo  {journal}
  {JCAP}\ }\textbf {\bibinfo {volume} {04}},\ \bibinfo {pages} {023} (\bibinfo
  {year} {2017}{\natexlab{a}})},\ \Eprint {http://arxiv.org/abs/1611.00376}
  {arXiv:1611.00376 [astro-ph.CO]} \BibitemShut {NoStop}%
\bibitem [{\citenamefont {Benisty}\ and\ \citenamefont
  {Staicova}(2021)}]{Benisty:2020otr}%
  \BibitemOpen
  \bibfield  {author} {\bibinfo {author} {\bibfnamefont {D.}~\bibnamefont
  {Benisty}}\ and\ \bibinfo {author} {\bibfnamefont {D.}~\bibnamefont
  {Staicova}},\ }\href {\doibase 10.1051/0004-6361/202039502} {\bibfield
  {journal} {\bibinfo  {journal} {Astron. Astrophys.}\ }\textbf {\bibinfo
  {volume} {647}},\ \bibinfo {pages} {A38} (\bibinfo {year} {2021})},\ \Eprint
  {http://arxiv.org/abs/2009.10701} {arXiv:2009.10701 [astro-ph.CO]}
  \BibitemShut {NoStop}%
\bibitem [{\citenamefont {Bernal}\ \emph {et~al.}(2016)\citenamefont {Bernal},
  \citenamefont {Verde},\ and\ \citenamefont {Riess}}]{Bernal:2016gxb}%
  \BibitemOpen
  \bibfield  {author} {\bibinfo {author} {\bibfnamefont {J.~L.}\ \bibnamefont
  {Bernal}}, \bibinfo {author} {\bibfnamefont {L.}~\bibnamefont {Verde}}, \
  and\ \bibinfo {author} {\bibfnamefont {A.~G.}\ \bibnamefont {Riess}},\ }\href
  {\doibase 10.1088/1475-7516/2016/10/019} {\bibfield  {journal} {\bibinfo
  {journal} {JCAP}\ }\textbf {\bibinfo {volume} {10}},\ \bibinfo {pages} {019}
  (\bibinfo {year} {2016})},\ \Eprint {http://arxiv.org/abs/1607.05617}
  {arXiv:1607.05617 [astro-ph.CO]} \BibitemShut {NoStop}%
\bibitem [{\citenamefont {Aylor}\ \emph {et~al.}(2019)\citenamefont {Aylor},
  \citenamefont {Joy}, \citenamefont {Knox}, \citenamefont {Millea},
  \citenamefont {Raghunathan},\ and\ \citenamefont {Wu}}]{Aylor:2018drw}%
  \BibitemOpen
  \bibfield  {author} {\bibinfo {author} {\bibfnamefont {K.}~\bibnamefont
  {Aylor}}, \bibinfo {author} {\bibfnamefont {M.}~\bibnamefont {Joy}}, \bibinfo
  {author} {\bibfnamefont {L.}~\bibnamefont {Knox}}, \bibinfo {author}
  {\bibfnamefont {M.}~\bibnamefont {Millea}}, \bibinfo {author} {\bibfnamefont
  {S.}~\bibnamefont {Raghunathan}}, \ and\ \bibinfo {author} {\bibfnamefont
  {W.~L.~K.}\ \bibnamefont {Wu}},\ }\href {\doibase 10.3847/1538-4357/ab0898}
  {\bibfield  {journal} {\bibinfo  {journal} {Astrophys. J.}\ }\textbf
  {\bibinfo {volume} {874}},\ \bibinfo {pages} {4} (\bibinfo {year} {2019})},\
  \Eprint {http://arxiv.org/abs/1811.00537} {arXiv:1811.00537 [astro-ph.CO]}
  \BibitemShut {NoStop}%
\bibitem [{\citenamefont {Heavens}\ \emph {et~al.}(2014)\citenamefont
  {Heavens}, \citenamefont {Jim\'enez},\ and\ \citenamefont
  {Verde}}]{Heavens:2014rja}%
  \BibitemOpen
  \bibfield  {author} {\bibinfo {author} {\bibfnamefont {A.}~\bibnamefont
  {Heavens}}, \bibinfo {author} {\bibfnamefont {R.}~\bibnamefont {Jim\'enez}},
  \ and\ \bibinfo {author} {\bibfnamefont {L.}~\bibnamefont {Verde}},\ }\href
  {\doibase 10.1103/PhysRevLett.113.241302} {\bibfield  {journal} {\bibinfo
  {journal} {Phys. Rev. Lett.}\ }\textbf {\bibinfo {volume} {113}},\ \bibinfo
  {pages} {241302} (\bibinfo {year} {2014})},\ \Eprint
  {http://arxiv.org/abs/1409.6217} {arXiv:1409.6217 [astro-ph.CO]} \BibitemShut
  {NoStop}%
\bibitem [{\citenamefont {Verde}\ \emph
  {et~al.}(2017{\natexlab{b}})\citenamefont {Verde}, \citenamefont {Bernal},
  \citenamefont {Heavens},\ and\ \citenamefont {Jim\'enez}}]{Verde:2016ccp}%
  \BibitemOpen
  \bibfield  {author} {\bibinfo {author} {\bibfnamefont {L.}~\bibnamefont
  {Verde}}, \bibinfo {author} {\bibfnamefont {J.~L.}\ \bibnamefont {Bernal}},
  \bibinfo {author} {\bibfnamefont {A.~F.}\ \bibnamefont {Heavens}}, \ and\
  \bibinfo {author} {\bibfnamefont {R.}~\bibnamefont {Jim\'enez}},\ }\href
  {\doibase 10.1093/mnras/stx116} {\bibfield  {journal} {\bibinfo  {journal}
  {Mon. Not. Roy. Astron. Soc.}\ }\textbf {\bibinfo {volume} {467}},\ \bibinfo
  {pages} {731} (\bibinfo {year} {2017}{\natexlab{b}})},\ \Eprint
  {http://arxiv.org/abs/1607.05297} {arXiv:1607.05297 [astro-ph.CO]}
  \BibitemShut {NoStop}%
\bibitem [{\citenamefont {Haridasu}\ \emph {et~al.}(2018)\citenamefont
  {Haridasu}, \citenamefont {Lukovi\'c}, \citenamefont {Moresco},\ and\
  \citenamefont {Vittorio}}]{Haridasu:2018gqm}%
  \BibitemOpen
  \bibfield  {author} {\bibinfo {author} {\bibfnamefont {B.~S.}\ \bibnamefont
  {Haridasu}}, \bibinfo {author} {\bibfnamefont {V.~V.}\ \bibnamefont
  {Lukovi\'c}}, \bibinfo {author} {\bibfnamefont {M.}~\bibnamefont {Moresco}},
  \ and\ \bibinfo {author} {\bibfnamefont {N.}~\bibnamefont {Vittorio}},\
  }\href {\doibase 10.1088/1475-7516/2018/10/015} {\bibfield  {journal}
  {\bibinfo  {journal} {JCAP}\ }\textbf {\bibinfo {volume} {10}},\ \bibinfo
  {pages} {015} (\bibinfo {year} {2018})},\ \Eprint
  {http://arxiv.org/abs/1805.03595} {arXiv:1805.03595 [astro-ph.CO]}
  \BibitemShut {NoStop}%
\bibitem [{\citenamefont {Cai}\ \emph {et~al.}(2022)\citenamefont {Cai},
  \citenamefont {Guo}, \citenamefont {Wang}, \citenamefont {Yu},\ and\
  \citenamefont {Zhou}}]{Cai:2021weh}%
  \BibitemOpen
  \bibfield  {author} {\bibinfo {author} {\bibfnamefont {R.-G.}\ \bibnamefont
  {Cai}}, \bibinfo {author} {\bibfnamefont {Z.-K.}\ \bibnamefont {Guo}},
  \bibinfo {author} {\bibfnamefont {S.-J.}\ \bibnamefont {Wang}}, \bibinfo
  {author} {\bibfnamefont {W.-W.}\ \bibnamefont {Yu}}, \ and\ \bibinfo {author}
  {\bibfnamefont {Y.}~\bibnamefont {Zhou}},\ }\href {\doibase
  10.1103/PhysRevD.105.L021301} {\bibfield  {journal} {\bibinfo  {journal}
  {Phys. Rev. D}\ }\textbf {\bibinfo {volume} {105}},\ \bibinfo {pages}
  {L021301} (\bibinfo {year} {2022})},\ \Eprint
  {http://arxiv.org/abs/2107.13286} {arXiv:2107.13286 [astro-ph.CO]}
  \BibitemShut {NoStop}%
\bibitem [{\citenamefont {L'Huillier}\ and\ \citenamefont
  {Shafieloo}(2017)}]{LHuillier:2016mtc}%
  \BibitemOpen
  \bibfield  {author} {\bibinfo {author} {\bibfnamefont {B.}~\bibnamefont
  {L'Huillier}}\ and\ \bibinfo {author} {\bibfnamefont {A.}~\bibnamefont
  {Shafieloo}},\ }\href {\doibase 10.1088/1475-7516/2017/01/015} {\bibfield
  {journal} {\bibinfo  {journal} {JCAP}\ }\textbf {\bibinfo {volume} {01}},\
  \bibinfo {pages} {015} (\bibinfo {year} {2017})},\ \Eprint
  {http://arxiv.org/abs/1606.06832} {arXiv:1606.06832 [astro-ph.CO]}
  \BibitemShut {NoStop}%
\bibitem [{\citenamefont {Shafieloo}\ \emph {et~al.}(2018)\citenamefont
  {Shafieloo}, \citenamefont {L'Huillier},\ and\ \citenamefont
  {Starobinsky}}]{Shafieloo:2018gin}%
  \BibitemOpen
  \bibfield  {author} {\bibinfo {author} {\bibfnamefont {A.}~\bibnamefont
  {Shafieloo}}, \bibinfo {author} {\bibfnamefont {B.}~\bibnamefont
  {L'Huillier}}, \ and\ \bibinfo {author} {\bibfnamefont {A.~A.}\ \bibnamefont
  {Starobinsky}},\ }\href {\doibase 10.1103/PhysRevD.98.083526} {\bibfield
  {journal} {\bibinfo  {journal} {Phys. Rev. D}\ }\textbf {\bibinfo {volume}
  {98}},\ \bibinfo {pages} {083526} (\bibinfo {year} {2018})},\ \Eprint
  {http://arxiv.org/abs/1804.04320} {arXiv:1804.04320 [astro-ph.CO]}
  \BibitemShut {NoStop}%
\bibitem [{\citenamefont {Lin}\ and\ \citenamefont
  {Ishak}(2017{\natexlab{a}})}]{Lin:2017ikq}%
  \BibitemOpen
  \bibfield  {author} {\bibinfo {author} {\bibfnamefont {W.}~\bibnamefont
  {Lin}}\ and\ \bibinfo {author} {\bibfnamefont {M.}~\bibnamefont {Ishak}},\
  }\href {\doibase 10.1103/PhysRevD.96.023532} {\bibfield  {journal} {\bibinfo
  {journal} {Phys. Rev. D}\ }\textbf {\bibinfo {volume} {96}},\ \bibinfo
  {pages} {023532} (\bibinfo {year} {2017}{\natexlab{a}})},\ \Eprint
  {http://arxiv.org/abs/1705.05303} {arXiv:1705.05303 [astro-ph.CO]}
  \BibitemShut {NoStop}%
\bibitem [{\citenamefont {Lin}\ and\ \citenamefont
  {Ishak}(2017{\natexlab{b}})}]{Lin:2017bhs}%
  \BibitemOpen
  \bibfield  {author} {\bibinfo {author} {\bibfnamefont {W.}~\bibnamefont
  {Lin}}\ and\ \bibinfo {author} {\bibfnamefont {M.}~\bibnamefont {Ishak}},\
  }\href {\doibase 10.1103/PhysRevD.96.083532} {\bibfield  {journal} {\bibinfo
  {journal} {Phys. Rev. D}\ }\textbf {\bibinfo {volume} {96}},\ \bibinfo
  {pages} {083532} (\bibinfo {year} {2017}{\natexlab{b}})},\ \Eprint
  {http://arxiv.org/abs/1708.09813} {arXiv:1708.09813 [astro-ph.CO]}
  \BibitemShut {NoStop}%
\bibitem [{\citenamefont {Garcia-Quintero}\ \emph {et~al.}(2019)\citenamefont
  {Garcia-Quintero}, \citenamefont {Ishak}, \citenamefont {Fox},\ and\
  \citenamefont {Lin}}]{Garcia-Quintero:2019cgt}%
  \BibitemOpen
  \bibfield  {author} {\bibinfo {author} {\bibfnamefont {C.}~\bibnamefont
  {Garcia-Quintero}}, \bibinfo {author} {\bibfnamefont {M.}~\bibnamefont
  {Ishak}}, \bibinfo {author} {\bibfnamefont {L.}~\bibnamefont {Fox}}, \ and\
  \bibinfo {author} {\bibfnamefont {W.}~\bibnamefont {Lin}},\ }\href {\doibase
  10.1103/PhysRevD.100.123538} {\bibfield  {journal} {\bibinfo  {journal}
  {Phys. Rev. D}\ }\textbf {\bibinfo {volume} {100}},\ \bibinfo {pages}
  {123538} (\bibinfo {year} {2019})},\ \Eprint
  {http://arxiv.org/abs/1910.01608} {arXiv:1910.01608 [astro-ph.CO]}
  \BibitemShut {NoStop}%
\bibitem [{\citenamefont {Gonzalez}\ \emph {et~al.}(2021)\citenamefont
  {Gonzalez}, \citenamefont {Benetti}, \citenamefont {von Marttens},\ and\
  \citenamefont {Alcaniz}}]{Gonzalez:2021ojp}%
  \BibitemOpen
  \bibfield  {author} {\bibinfo {author} {\bibfnamefont {J.~E.}\ \bibnamefont
  {Gonzalez}}, \bibinfo {author} {\bibfnamefont {M.}~\bibnamefont {Benetti}},
  \bibinfo {author} {\bibfnamefont {R.}~\bibnamefont {von Marttens}}, \ and\
  \bibinfo {author} {\bibfnamefont {J.}~\bibnamefont {Alcaniz}},\ }\href
  {\doibase 10.1088/1475-7516/2021/11/060} {\bibfield  {journal} {\bibinfo
  {journal} {JCAP}\ }\textbf {\bibinfo {volume} {11}},\ \bibinfo {pages} {060}
  (\bibinfo {year} {2021})},\ \Eprint {http://arxiv.org/abs/2104.13455}
  {arXiv:2104.13455 [astro-ph.CO]} \BibitemShut {NoStop}%
\bibitem [{\citenamefont {Scolnic}\ \emph {et~al.}(2018)\citenamefont {Scolnic}
  \emph {et~al.}}]{Scolnic:2017caz}%
  \BibitemOpen
  \bibfield  {author} {\bibinfo {author} {\bibfnamefont {D.~M.}\ \bibnamefont
  {Scolnic}} \emph {et~al.},\ }\href {\doibase 10.3847/1538-4357/aab9bb}
  {\bibfield  {journal} {\bibinfo  {journal} {Astrophys. J.}\ }\textbf
  {\bibinfo {volume} {859}},\ \bibinfo {pages} {101} (\bibinfo {year}
  {2018})},\ \Eprint {http://arxiv.org/abs/1710.00845} {arXiv:1710.00845
  [astro-ph.CO]} \BibitemShut {NoStop}%
\bibitem [{\citenamefont {Cole}\ \emph {et~al.}(2005)\citenamefont {Cole} \emph
  {et~al.}}]{Cole:2005sx}%
  \BibitemOpen
  \bibfield  {author} {\bibinfo {author} {\bibfnamefont {S.}~\bibnamefont
  {Cole}} \emph {et~al.} (\bibinfo {collaboration} {2dFGRS}),\ }\href {\doibase
  10.1111/j.1365-2966.2005.09318.x} {\bibfield  {journal} {\bibinfo  {journal}
  {Mon. Not. Roy. Astron. Soc.}\ }\textbf {\bibinfo {volume} {362}},\ \bibinfo
  {pages} {505} (\bibinfo {year} {2005})},\ \Eprint
  {http://arxiv.org/abs/astro-ph/0501174} {arXiv:astro-ph/0501174 [astro-ph]}
  \BibitemShut {NoStop}%
\bibitem [{\citenamefont {Eisenstein}\ \emph {et~al.}(2005)\citenamefont
  {Eisenstein} \emph {et~al.}}]{Eisenstein:2005su}%
  \BibitemOpen
  \bibfield  {author} {\bibinfo {author} {\bibfnamefont {D.~J.}\ \bibnamefont
  {Eisenstein}} \emph {et~al.} (\bibinfo {collaboration} {SDSS}),\ }\href
  {\doibase 10.1086/466512} {\bibfield  {journal} {\bibinfo  {journal}
  {Astrophys. J.}\ }\textbf {\bibinfo {volume} {633}},\ \bibinfo {pages} {560}
  (\bibinfo {year} {2005})},\ \Eprint {http://arxiv.org/abs/astro-ph/0501171}
  {arXiv:astro-ph/0501171 [astro-ph]} \BibitemShut {NoStop}%
\bibitem [{\citenamefont {Bernal}\ \emph {et~al.}(2020)\citenamefont {Bernal},
  \citenamefont {Smith}, \citenamefont {Boddy},\ and\ \citenamefont
  {Kamionkowski}}]{Bernal:2020vbb}%
  \BibitemOpen
  \bibfield  {author} {\bibinfo {author} {\bibfnamefont {J.~L.}\ \bibnamefont
  {Bernal}}, \bibinfo {author} {\bibfnamefont {T.~L.}\ \bibnamefont {Smith}},
  \bibinfo {author} {\bibfnamefont {K.~K.}\ \bibnamefont {Boddy}}, \ and\
  \bibinfo {author} {\bibfnamefont {M.}~\bibnamefont {Kamionkowski}},\ }\href
  {\doibase 10.1103/PhysRevD.102.123515} {\bibfield  {journal} {\bibinfo
  {journal} {Phys. Rev. D}\ }\textbf {\bibinfo {volume} {102}},\ \bibinfo
  {pages} {123515} (\bibinfo {year} {2020})},\ \Eprint
  {http://arxiv.org/abs/2004.07263} {arXiv:2004.07263 [astro-ph.CO]}
  \BibitemShut {NoStop}%
\bibitem [{\citenamefont {Etherington}(1933)}]{Etherington:1933}%
  \BibitemOpen
  \bibfield  {author} {\bibinfo {author} {\bibfnamefont {I.}~\bibnamefont
  {Etherington}},\ }\href {\doibase 10.1080/14786443309462220} {\bibfield
  {journal} {\bibinfo  {journal} {Philos. Mag.}\ }\textbf {\bibinfo {volume}
  {15}},\ \bibinfo {pages} {761} (\bibinfo {year} {1933})}\BibitemShut
  {NoStop}%
\bibitem [{\citenamefont {Carter}\ \emph {et~al.}(2018)\citenamefont {Carter},
  \citenamefont {Beutler}, \citenamefont {Percival}, \citenamefont {Blake},
  \citenamefont {Koda},\ and\ \citenamefont {Ross}}]{Carter:2018vce}%
  \BibitemOpen
  \bibfield  {author} {\bibinfo {author} {\bibfnamefont {P.}~\bibnamefont
  {Carter}}, \bibinfo {author} {\bibfnamefont {F.}~\bibnamefont {Beutler}},
  \bibinfo {author} {\bibfnamefont {W.~J.}\ \bibnamefont {Percival}}, \bibinfo
  {author} {\bibfnamefont {C.}~\bibnamefont {Blake}}, \bibinfo {author}
  {\bibfnamefont {J.}~\bibnamefont {Koda}}, \ and\ \bibinfo {author}
  {\bibfnamefont {A.~J.}\ \bibnamefont {Ross}},\ }\href {\doibase
  10.1093/mnras/sty2405} {\bibfield  {journal} {\bibinfo  {journal} {Mon. Not.
  Roy. Astron. Soc.}\ }\textbf {\bibinfo {volume} {481}},\ \bibinfo {pages}
  {2371} (\bibinfo {year} {2018})},\ \Eprint {http://arxiv.org/abs/1803.01746}
  {arXiv:1803.01746 [astro-ph.CO]} \BibitemShut {NoStop}%
\bibitem [{\citenamefont {Beutler}\ \emph {et~al.}(2011)\citenamefont
  {Beutler}, \citenamefont {Blake}, \citenamefont {Colless}, \citenamefont
  {Jones}, \citenamefont {Staveley-Smith}, \citenamefont {Campbell},
  \citenamefont {Parker}, \citenamefont {Saunders},\ and\ \citenamefont
  {Watson}}]{Beutler:2011hx}%
  \BibitemOpen
  \bibfield  {author} {\bibinfo {author} {\bibfnamefont {F.}~\bibnamefont
  {Beutler}}, \bibinfo {author} {\bibfnamefont {C.}~\bibnamefont {Blake}},
  \bibinfo {author} {\bibfnamefont {M.}~\bibnamefont {Colless}}, \bibinfo
  {author} {\bibfnamefont {D.~H.}\ \bibnamefont {Jones}}, \bibinfo {author}
  {\bibfnamefont {L.}~\bibnamefont {Staveley-Smith}}, \bibinfo {author}
  {\bibfnamefont {L.}~\bibnamefont {Campbell}}, \bibinfo {author}
  {\bibfnamefont {Q.}~\bibnamefont {Parker}}, \bibinfo {author} {\bibfnamefont
  {W.}~\bibnamefont {Saunders}}, \ and\ \bibinfo {author} {\bibfnamefont
  {F.}~\bibnamefont {Watson}},\ }\href {\doibase
  10.1111/j.1365-2966.2011.19250.x} {\bibfield  {journal} {\bibinfo  {journal}
  {Mon. Not. Roy. Astron. Soc.}\ }\textbf {\bibinfo {volume} {416}},\ \bibinfo
  {pages} {3017} (\bibinfo {year} {2011})},\ \Eprint
  {http://arxiv.org/abs/1106.3366} {arXiv:1106.3366 [astro-ph.CO]} \BibitemShut
  {NoStop}%
\bibitem [{\citenamefont {Ross}\ \emph {et~al.}(2015)\citenamefont {Ross},
  \citenamefont {Samushia}, \citenamefont {Howlett}, \citenamefont {Percival},
  \citenamefont {Burden},\ and\ \citenamefont {Manera}}]{Ross:2014qpa}%
  \BibitemOpen
  \bibfield  {author} {\bibinfo {author} {\bibfnamefont {A.~J.}\ \bibnamefont
  {Ross}}, \bibinfo {author} {\bibfnamefont {L.}~\bibnamefont {Samushia}},
  \bibinfo {author} {\bibfnamefont {C.}~\bibnamefont {Howlett}}, \bibinfo
  {author} {\bibfnamefont {W.~J.}\ \bibnamefont {Percival}}, \bibinfo {author}
  {\bibfnamefont {A.}~\bibnamefont {Burden}}, \ and\ \bibinfo {author}
  {\bibfnamefont {M.}~\bibnamefont {Manera}},\ }\href {\doibase
  10.1093/mnras/stv154} {\bibfield  {journal} {\bibinfo  {journal} {Mon. Not.
  Roy. Astron. Soc.}\ }\textbf {\bibinfo {volume} {449}},\ \bibinfo {pages}
  {835} (\bibinfo {year} {2015})},\ \Eprint {http://arxiv.org/abs/1409.3242}
  {arXiv:1409.3242 [astro-ph.CO]} \BibitemShut {NoStop}%
\bibitem [{\citenamefont {Gil-Mar\'\i{}n}\ \emph {et~al.}(2017)\citenamefont
  {Gil-Mar\'\i{}n}, \citenamefont {Percival}, \citenamefont {Verde},
  \citenamefont {Brownstein}, \citenamefont {Chuang}, \citenamefont {Kitaura},
  \citenamefont {Rodr\'\i{}guez-Torres},\ and\ \citenamefont
  {Olmstead}}]{Gil-Marin:2016wya}%
  \BibitemOpen
  \bibfield  {author} {\bibinfo {author} {\bibfnamefont {H.}~\bibnamefont
  {Gil-Mar\'\i{}n}}, \bibinfo {author} {\bibfnamefont {W.~J.}\ \bibnamefont
  {Percival}}, \bibinfo {author} {\bibfnamefont {L.}~\bibnamefont {Verde}},
  \bibinfo {author} {\bibfnamefont {J.~R.}\ \bibnamefont {Brownstein}},
  \bibinfo {author} {\bibfnamefont {C.-H.}\ \bibnamefont {Chuang}}, \bibinfo
  {author} {\bibfnamefont {F.-S.}\ \bibnamefont {Kitaura}}, \bibinfo {author}
  {\bibfnamefont {S.~A.}\ \bibnamefont {Rodr\'\i{}guez-Torres}}, \ and\
  \bibinfo {author} {\bibfnamefont {M.~D.}\ \bibnamefont {Olmstead}},\ }\href
  {\doibase 10.1093/mnras/stw2679} {\bibfield  {journal} {\bibinfo  {journal}
  {Mon. Not. Roy. Astron. Soc.}\ }\textbf {\bibinfo {volume} {465}},\ \bibinfo
  {pages} {1757} (\bibinfo {year} {2017})},\ \Eprint
  {http://arxiv.org/abs/1606.00439} {arXiv:1606.00439 [astro-ph.CO]}
  \BibitemShut {NoStop}%
\bibitem [{\citenamefont {Kazin}\ \emph {et~al.}(2014)\citenamefont {Kazin}
  \emph {et~al.}}]{Kazin:2014qga}%
  \BibitemOpen
  \bibfield  {author} {\bibinfo {author} {\bibfnamefont {E.~A.}\ \bibnamefont
  {Kazin}} \emph {et~al.},\ }\href {\doibase 10.1093/mnras/stu778} {\bibfield
  {journal} {\bibinfo  {journal} {Mon. Not. Roy. Astron. Soc.}\ }\textbf
  {\bibinfo {volume} {441}},\ \bibinfo {pages} {3524} (\bibinfo {year}
  {2014})},\ \Eprint {http://arxiv.org/abs/1401.0358} {arXiv:1401.0358
  [astro-ph.CO]} \BibitemShut {NoStop}%
\bibitem [{\citenamefont {Abbott}\ \emph {et~al.}(2019)\citenamefont {Abbott}
  \emph {et~al.}}]{DES:2017rfo}%
  \BibitemOpen
  \bibfield  {author} {\bibinfo {author} {\bibfnamefont {T.~M.~C.}\
  \bibnamefont {Abbott}} \emph {et~al.} (\bibinfo {collaboration} {DES}),\
  }\href {\doibase 10.1093/mnras/sty3351} {\bibfield  {journal} {\bibinfo
  {journal} {Mon. Not. Roy. Astron. Soc.}\ }\textbf {\bibinfo {volume} {483}},\
  \bibinfo {pages} {4866} (\bibinfo {year} {2019})},\ \Eprint
  {http://arxiv.org/abs/1712.06209} {arXiv:1712.06209 [astro-ph.CO]}
  \BibitemShut {NoStop}%
\bibitem [{\citenamefont {Abbott}\ \emph {et~al.}(2021)\citenamefont {Abbott}
  \emph {et~al.}}]{DES:2021esc}%
  \BibitemOpen
  \bibfield  {author} {\bibinfo {author} {\bibfnamefont {T.~M.~C.}\
  \bibnamefont {Abbott}} \emph {et~al.} (\bibinfo {collaboration} {DES}),\
  }\href@noop {} {\  (\bibinfo {year} {2021})},\ \Eprint
  {http://arxiv.org/abs/2107.04646} {arXiv:2107.04646 [astro-ph.CO]}
  \BibitemShut {NoStop}%
\bibitem [{\citenamefont {Neveux}\ \emph {et~al.}(2020)\citenamefont {Neveux}
  \emph {et~al.}}]{Neveux:2020voa}%
  \BibitemOpen
  \bibfield  {author} {\bibinfo {author} {\bibfnamefont {R.}~\bibnamefont
  {Neveux}} \emph {et~al.},\ }\href {\doibase 10.1093/mnras/staa2780}
  {\bibfield  {journal} {\bibinfo  {journal} {Mon. Not. Roy. Astron. Soc.}\
  }\textbf {\bibinfo {volume} {499}},\ \bibinfo {pages} {210} (\bibinfo {year}
  {2020})},\ \Eprint {http://arxiv.org/abs/2007.08999} {arXiv:2007.08999
  [astro-ph.CO]} \BibitemShut {NoStop}%
\bibitem [{\citenamefont {Hou}\ \emph {et~al.}(2020)\citenamefont {Hou} \emph
  {et~al.}}]{Hou:2020rse}%
  \BibitemOpen
  \bibfield  {author} {\bibinfo {author} {\bibfnamefont {J.}~\bibnamefont
  {Hou}} \emph {et~al.},\ }\href {\doibase 10.1093/mnras/staa3234} {\bibfield
  {journal} {\bibinfo  {journal} {Mon. Not. Roy. Astron. Soc.}\ }\textbf
  {\bibinfo {volume} {500}},\ \bibinfo {pages} {1201} (\bibinfo {year}
  {2020})},\ \Eprint {http://arxiv.org/abs/2007.08998} {arXiv:2007.08998
  [astro-ph.CO]} \BibitemShut {NoStop}%
\bibitem [{\citenamefont {du~Mas~des Bourboux}\ \emph
  {et~al.}(2020)\citenamefont {du~Mas~des Bourboux} \emph
  {et~al.}}]{duMasdesBourboux:2020pck}%
  \BibitemOpen
  \bibfield  {author} {\bibinfo {author} {\bibfnamefont {H.}~\bibnamefont
  {du~Mas~des Bourboux}} \emph {et~al.},\ }\href {\doibase
  10.3847/1538-4357/abb085} {\bibfield  {journal} {\bibinfo  {journal}
  {Astrophys. J.}\ }\textbf {\bibinfo {volume} {901}},\ \bibinfo {pages} {153}
  (\bibinfo {year} {2020})},\ \Eprint {http://arxiv.org/abs/2007.08995}
  {arXiv:2007.08995 [astro-ph.CO]} \BibitemShut {NoStop}%
\bibitem [{\citenamefont {Jim{\'{e}}nez}\ and\ \citenamefont
  {Loeb}(2002)}]{Jimenez:2001gg}%
  \BibitemOpen
  \bibfield  {author} {\bibinfo {author} {\bibfnamefont {R.}~\bibnamefont
  {Jim{\'{e}}nez}}\ and\ \bibinfo {author} {\bibfnamefont {A.}~\bibnamefont
  {Loeb}},\ }\href {\doibase 10.1086/340549} {\bibfield  {journal} {\bibinfo
  {journal} {Astrophys. J.}\ }\textbf {\bibinfo {volume} {573}},\ \bibinfo
  {pages} {37} (\bibinfo {year} {2002})},\ \Eprint
  {http://arxiv.org/abs/astro-ph/0106145} {arXiv:astro-ph/0106145 [astro-ph]}
  \BibitemShut {NoStop}%
\bibitem [{\citenamefont {Moresco}\ \emph {et~al.}(2020)\citenamefont
  {Moresco}, \citenamefont {Jim\'enez}, \citenamefont {Verde}, \citenamefont
  {Cimatti},\ and\ \citenamefont {Pozzetti}}]{Moresco:2020fbm}%
  \BibitemOpen
  \bibfield  {author} {\bibinfo {author} {\bibfnamefont {M.}~\bibnamefont
  {Moresco}}, \bibinfo {author} {\bibfnamefont {R.}~\bibnamefont {Jim\'enez}},
  \bibinfo {author} {\bibfnamefont {L.}~\bibnamefont {Verde}}, \bibinfo
  {author} {\bibfnamefont {A.}~\bibnamefont {Cimatti}}, \ and\ \bibinfo
  {author} {\bibfnamefont {L.}~\bibnamefont {Pozzetti}},\ }\href {\doibase
  10.3847/1538-4357/ab9eb0} {\bibfield  {journal} {\bibinfo  {journal}
  {Astrophys. J.}\ }\textbf {\bibinfo {volume} {898}},\ \bibinfo {pages} {82}
  (\bibinfo {year} {2020})},\ \Eprint {http://arxiv.org/abs/2003.07362}
  {arXiv:2003.07362 [astro-ph.GA]} \BibitemShut {NoStop}%
\bibitem [{\citenamefont {L\'opez-Corredoira}\ \emph
  {et~al.}(2017)\citenamefont {L\'opez-Corredoira}, \citenamefont {Vazdekis},
  \citenamefont {Guti\'errez},\ and\ \citenamefont
  {Castro-Rodr\'iguez}}]{Lopez-Corredoira:2017zfl}%
  \BibitemOpen
  \bibfield  {author} {\bibinfo {author} {\bibfnamefont {M.}~\bibnamefont
  {L\'opez-Corredoira}}, \bibinfo {author} {\bibfnamefont {A.}~\bibnamefont
  {Vazdekis}}, \bibinfo {author} {\bibfnamefont {C.~M.}\ \bibnamefont
  {Guti\'errez}}, \ and\ \bibinfo {author} {\bibfnamefont {N.}~\bibnamefont
  {Castro-Rodr\'iguez}},\ }\href {\doibase 10.1051/0004-6361/201629857}
  {\bibfield  {journal} {\bibinfo  {journal} {Astron. Astrophys.}\ }\textbf
  {\bibinfo {volume} {600}},\ \bibinfo {pages} {A91} (\bibinfo {year}
  {2017})},\ \Eprint {http://arxiv.org/abs/1702.00380} {arXiv:1702.00380
  [astro-ph.GA]} \BibitemShut {NoStop}%
\bibitem [{\citenamefont {L\'opez-Corredoira}\ and\ \citenamefont
  {Vazdekis}(2018)}]{Lopez-Corredoira:2018tmn}%
  \BibitemOpen
  \bibfield  {author} {\bibinfo {author} {\bibfnamefont {M.}~\bibnamefont
  {L\'opez-Corredoira}}\ and\ \bibinfo {author} {\bibfnamefont
  {A.}~\bibnamefont {Vazdekis}},\ }\href {\doibase 10.1051/0004-6361/201731647}
  {\bibfield  {journal} {\bibinfo  {journal} {Astron. Astrophys.}\ }\textbf
  {\bibinfo {volume} {614}},\ \bibinfo {pages} {A127} (\bibinfo {year}
  {2018})},\ \Eprint {http://arxiv.org/abs/1802.09473} {arXiv:1802.09473
  [astro-ph.CO]} \BibitemShut {NoStop}%
\bibitem [{\citenamefont {Moresco}\ \emph {et~al.}(2018)\citenamefont
  {Moresco}, \citenamefont {Jim\'enez}, \citenamefont {Verde}, \citenamefont
  {Pozzetti}, \citenamefont {Cimatti},\ and\ \citenamefont
  {Citro}}]{Moresco:2018xdr}%
  \BibitemOpen
  \bibfield  {author} {\bibinfo {author} {\bibfnamefont {M.}~\bibnamefont
  {Moresco}}, \bibinfo {author} {\bibfnamefont {R.}~\bibnamefont {Jim\'enez}},
  \bibinfo {author} {\bibfnamefont {L.}~\bibnamefont {Verde}}, \bibinfo
  {author} {\bibfnamefont {L.}~\bibnamefont {Pozzetti}}, \bibinfo {author}
  {\bibfnamefont {A.}~\bibnamefont {Cimatti}}, \ and\ \bibinfo {author}
  {\bibfnamefont {A.}~\bibnamefont {Citro}},\ }\href {\doibase
  10.3847/1538-4357/aae829} {\bibfield  {journal} {\bibinfo  {journal}
  {Astrophys. J.}\ }\textbf {\bibinfo {volume} {868}},\ \bibinfo {pages} {84}
  (\bibinfo {year} {2018})},\ \Eprint {http://arxiv.org/abs/1804.05864}
  {arXiv:1804.05864 [astro-ph.CO]} \BibitemShut {NoStop}%
\bibitem [{\citenamefont {Jim{\'{e}}nez}\ \emph {et~al.}(2003)\citenamefont
  {Jim{\'{e}}nez}, \citenamefont {Verde}, \citenamefont {Treu},\ and\
  \citenamefont {Stern}}]{Jimenez:2003iv}%
  \BibitemOpen
  \bibfield  {author} {\bibinfo {author} {\bibfnamefont {R.}~\bibnamefont
  {Jim{\'{e}}nez}}, \bibinfo {author} {\bibfnamefont {L.}~\bibnamefont
  {Verde}}, \bibinfo {author} {\bibfnamefont {T.}~\bibnamefont {Treu}}, \ and\
  \bibinfo {author} {\bibfnamefont {D.}~\bibnamefont {Stern}},\ }\href
  {\doibase 10.1086/376595} {\bibfield  {journal} {\bibinfo  {journal}
  {Astrophys. J.}\ }\textbf {\bibinfo {volume} {593}},\ \bibinfo {pages} {622}
  (\bibinfo {year} {2003})},\ \Eprint {http://arxiv.org/abs/astro-ph/0302560}
  {arXiv:astro-ph/0302560 [astro-ph]} \BibitemShut {NoStop}%
\bibitem [{\citenamefont {Stern}\ \emph {et~al.}(2010)\citenamefont {Stern},
  \citenamefont {Jim{\'{e}}nez}, \citenamefont {Verde}, \citenamefont
  {Kamionkowski},\ and\ \citenamefont {Stanford}}]{Stern:2009ep}%
  \BibitemOpen
  \bibfield  {author} {\bibinfo {author} {\bibfnamefont {D.}~\bibnamefont
  {Stern}}, \bibinfo {author} {\bibfnamefont {R.}~\bibnamefont
  {Jim{\'{e}}nez}}, \bibinfo {author} {\bibfnamefont {L.}~\bibnamefont
  {Verde}}, \bibinfo {author} {\bibfnamefont {M.}~\bibnamefont {Kamionkowski}},
  \ and\ \bibinfo {author} {\bibfnamefont {S.~A.}\ \bibnamefont {Stanford}},\
  }\href {\doibase 10.1088/1475-7516/2010/02/008} {\bibfield  {journal}
  {\bibinfo  {journal} {JCAP}\ }\textbf {\bibinfo {volume} {1002}},\ \bibinfo
  {pages} {008} (\bibinfo {year} {2010})},\ \Eprint
  {http://arxiv.org/abs/0907.3149} {arXiv:0907.3149 [astro-ph.CO]} \BibitemShut
  {NoStop}%
\bibitem [{\citenamefont {Moresco}\ \emph {et~al.}(2012)\citenamefont {Moresco}
  \emph {et~al.}}]{Moresco:2012jh}%
  \BibitemOpen
  \bibfield  {author} {\bibinfo {author} {\bibfnamefont {M.}~\bibnamefont
  {Moresco}} \emph {et~al.},\ }\href {\doibase 10.1088/1475-7516/2012/08/006}
  {\bibfield  {journal} {\bibinfo  {journal} {JCAP}\ }\textbf {\bibinfo
  {volume} {1208}},\ \bibinfo {pages} {006} (\bibinfo {year} {2012})},\ \Eprint
  {http://arxiv.org/abs/1201.3609} {arXiv:1201.3609 [astro-ph.CO]} \BibitemShut
  {NoStop}%
\bibitem [{\citenamefont {Zhang}\ \emph {et~al.}(2014)\citenamefont {Zhang},
  \citenamefont {Zhang}, \citenamefont {Yuan}, \citenamefont {Zhang},\ and\
  \citenamefont {Sun}}]{Zhang:2012mp}%
  \BibitemOpen
  \bibfield  {author} {\bibinfo {author} {\bibfnamefont {C.}~\bibnamefont
  {Zhang}}, \bibinfo {author} {\bibfnamefont {H.}~\bibnamefont {Zhang}},
  \bibinfo {author} {\bibfnamefont {S.}~\bibnamefont {Yuan}}, \bibinfo {author}
  {\bibfnamefont {T.-J.}\ \bibnamefont {Zhang}}, \ and\ \bibinfo {author}
  {\bibfnamefont {Y.-C.}\ \bibnamefont {Sun}},\ }\href {\doibase
  10.1088/1674-4527/14/10/002} {\bibfield  {journal} {\bibinfo  {journal} {Res.
  Astron. Astrophys.}\ }\textbf {\bibinfo {volume} {14}},\ \bibinfo {pages}
  {1221} (\bibinfo {year} {2014})},\ \Eprint {http://arxiv.org/abs/1207.4541}
  {arXiv:1207.4541 [astro-ph.CO]} \BibitemShut {NoStop}%
\bibitem [{\citenamefont {Moresco}(2015)}]{Moresco:2015cya}%
  \BibitemOpen
  \bibfield  {author} {\bibinfo {author} {\bibfnamefont {M.}~\bibnamefont
  {Moresco}},\ }\href {\doibase 10.1093/mnrasl/slv037} {\bibfield  {journal}
  {\bibinfo  {journal} {Mon. Not. Roy. Astron. Soc.}\ }\textbf {\bibinfo
  {volume} {450}},\ \bibinfo {pages} {L16} (\bibinfo {year} {2015})},\ \Eprint
  {http://arxiv.org/abs/1503.01116} {arXiv:1503.01116 [astro-ph.CO]}
  \BibitemShut {NoStop}%
\bibitem [{\citenamefont {Moresco}\ \emph {et~al.}(2016)\citenamefont
  {Moresco}, \citenamefont {Pozzetti}, \citenamefont {Cimatti}, \citenamefont
  {Jim{\'{e}}nez}, \citenamefont {Maraston}, \citenamefont {Verde},
  \citenamefont {Thomas}, \citenamefont {Citro}, \citenamefont {Tojeiro},\ and\
  \citenamefont {Wilkinson}}]{Moresco:2016mzx}%
  \BibitemOpen
  \bibfield  {author} {\bibinfo {author} {\bibfnamefont {M.}~\bibnamefont
  {Moresco}}, \bibinfo {author} {\bibfnamefont {L.}~\bibnamefont {Pozzetti}},
  \bibinfo {author} {\bibfnamefont {A.}~\bibnamefont {Cimatti}}, \bibinfo
  {author} {\bibfnamefont {R.}~\bibnamefont {Jim{\'{e}}nez}}, \bibinfo {author}
  {\bibfnamefont {C.}~\bibnamefont {Maraston}}, \bibinfo {author}
  {\bibfnamefont {L.}~\bibnamefont {Verde}}, \bibinfo {author} {\bibfnamefont
  {D.}~\bibnamefont {Thomas}}, \bibinfo {author} {\bibfnamefont
  {A.}~\bibnamefont {Citro}}, \bibinfo {author} {\bibfnamefont
  {R.}~\bibnamefont {Tojeiro}}, \ and\ \bibinfo {author} {\bibfnamefont
  {D.}~\bibnamefont {Wilkinson}},\ }\href {\doibase
  10.1088/1475-7516/2016/05/014} {\bibfield  {journal} {\bibinfo  {journal}
  {JCAP}\ }\textbf {\bibinfo {volume} {1605}},\ \bibinfo {pages} {014}
  (\bibinfo {year} {2016})},\ \Eprint {http://arxiv.org/abs/1601.01701}
  {arXiv:1601.01701 [astro-ph.CO]} \BibitemShut {NoStop}%
\bibitem [{\citenamefont {Ratsimbazafy}\ \emph {et~al.}(2017)\citenamefont
  {Ratsimbazafy}, \citenamefont {Loubser}, \citenamefont {Crawford},
  \citenamefont {Cress}, \citenamefont {Bassett}, \citenamefont {Nichol},\ and\
  \citenamefont {V{\"{a}}is{\"{a}}nen}}]{Ratsimbazafy:2017vga}%
  \BibitemOpen
  \bibfield  {author} {\bibinfo {author} {\bibfnamefont {A.~L.}\ \bibnamefont
  {Ratsimbazafy}}, \bibinfo {author} {\bibfnamefont {S.~I.}\ \bibnamefont
  {Loubser}}, \bibinfo {author} {\bibfnamefont {S.~M.}\ \bibnamefont
  {Crawford}}, \bibinfo {author} {\bibfnamefont {C.~M.}\ \bibnamefont {Cress}},
  \bibinfo {author} {\bibfnamefont {B.~A.}\ \bibnamefont {Bassett}}, \bibinfo
  {author} {\bibfnamefont {R.~C.}\ \bibnamefont {Nichol}}, \ and\ \bibinfo
  {author} {\bibfnamefont {P.}~\bibnamefont {V{\"{a}}is{\"{a}}nen}},\ }\href
  {\doibase 10.1093/mnras/stx301} {\bibfield  {journal} {\bibinfo  {journal}
  {Mon. Not. Roy. Astron. Soc.}\ }\textbf {\bibinfo {volume} {467}},\ \bibinfo
  {pages} {3239} (\bibinfo {year} {2017})},\ \Eprint
  {http://arxiv.org/abs/1702.00418} {arXiv:1702.00418 [astro-ph.CO]}
  \BibitemShut {NoStop}%
\bibitem [{\citenamefont {Borghi}\ \emph {et~al.}(2021)\citenamefont {Borghi},
  \citenamefont {Moresco},\ and\ \citenamefont {Cimatti}}]{Borghi:2021rft}%
  \BibitemOpen
  \bibfield  {author} {\bibinfo {author} {\bibfnamefont {N.}~\bibnamefont
  {Borghi}}, \bibinfo {author} {\bibfnamefont {M.}~\bibnamefont {Moresco}}, \
  and\ \bibinfo {author} {\bibfnamefont {A.}~\bibnamefont {Cimatti}},\
  }\href@noop {} {\  (\bibinfo {year} {2021})},\ \Eprint
  {http://arxiv.org/abs/2110.04304} {arXiv:2110.04304 [astro-ph.CO]}
  \BibitemShut {NoStop}%
\bibitem [{\citenamefont {G\'omez-Valent}(2019)}]{Gomez-Valent:2018gvm}%
  \BibitemOpen
  \bibfield  {author} {\bibinfo {author} {\bibfnamefont {A.}~\bibnamefont
  {G\'omez-Valent}},\ }\href {\doibase 10.1088/1475-7516/2019/05/026}
  {\bibfield  {journal} {\bibinfo  {journal} {JCAP}\ }\textbf {\bibinfo
  {volume} {05}},\ \bibinfo {pages} {026} (\bibinfo {year} {2019})},\ \Eprint
  {http://arxiv.org/abs/1810.02278} {arXiv:1810.02278 [astro-ph.CO]}
  \BibitemShut {NoStop}%
\bibitem [{\citenamefont {Simon}\ \emph {et~al.}(2005)\citenamefont {Simon},
  \citenamefont {Verde},\ and\ \citenamefont {Jim{\'{e}}nez}}]{Simon:2004tf}%
  \BibitemOpen
  \bibfield  {author} {\bibinfo {author} {\bibfnamefont {J.}~\bibnamefont
  {Simon}}, \bibinfo {author} {\bibfnamefont {L.}~\bibnamefont {Verde}}, \ and\
  \bibinfo {author} {\bibfnamefont {R.}~\bibnamefont {Jim{\'{e}}nez}},\ }\href
  {\doibase 10.1103/PhysRevD.71.123001} {\bibfield  {journal} {\bibinfo
  {journal} {Phys. Rev.}\ }\textbf {\bibinfo {volume} {D71}},\ \bibinfo {pages}
  {123001} (\bibinfo {year} {2005})},\ \Eprint
  {http://arxiv.org/abs/astro-ph/0412269} {arXiv:astro-ph/0412269 [astro-ph]}
  \BibitemShut {NoStop}%
\bibitem [{\citenamefont {Kjerrgren}\ and\ \citenamefont
  {Mortsell}(2021)}]{Kjerrgren:2021zuo}%
  \BibitemOpen
  \bibfield  {author} {\bibinfo {author} {\bibfnamefont {A.~A.}\ \bibnamefont
  {Kjerrgren}}\ and\ \bibinfo {author} {\bibfnamefont {E.}~\bibnamefont
  {Mortsell}},\ }\href@noop {} {\  (\bibinfo {year} {2021})},\ \Eprint
  {http://arxiv.org/abs/2106.11317} {arXiv:2106.11317 [astro-ph.CO]}
  \BibitemShut {NoStop}%
\bibitem [{\citenamefont {Mahalanobis}(1936)}]{Mahalanobis:1936}%
  \BibitemOpen
  \bibfield  {author} {\bibinfo {author} {\bibfnamefont {P.}~\bibnamefont
  {Mahalanobis}},\ }\href@noop {} {\bibfield  {journal} {\bibinfo  {journal}
  {Proceedings of the National Institute of Science of India}\ }\textbf
  {\bibinfo {volume} {2}},\ \bibinfo {pages} {49} (\bibinfo {year}
  {1936})}\BibitemShut {NoStop}%
\bibitem [{\citenamefont {Handley}(2021)}]{Handley:2019tkm}%
  \BibitemOpen
  \bibfield  {author} {\bibinfo {author} {\bibfnamefont {W.}~\bibnamefont
  {Handley}},\ }\href {\doibase 10.1103/PhysRevD.103.L041301} {\bibfield
  {journal} {\bibinfo  {journal} {Phys. Rev. D}\ }\textbf {\bibinfo {volume}
  {103}},\ \bibinfo {pages} {L041301} (\bibinfo {year} {2021})},\ \Eprint
  {http://arxiv.org/abs/1908.09139} {arXiv:1908.09139 [astro-ph.CO]}
  \BibitemShut {NoStop}%
\bibitem [{\citenamefont {Di~Valentino}\ \emph {et~al.}(2019)\citenamefont
  {Di~Valentino}, \citenamefont {Melchiorri},\ and\ \citenamefont
  {Silk}}]{DiValentino:2019qzk}%
  \BibitemOpen
  \bibfield  {author} {\bibinfo {author} {\bibfnamefont {E.}~\bibnamefont
  {Di~Valentino}}, \bibinfo {author} {\bibfnamefont {A.}~\bibnamefont
  {Melchiorri}}, \ and\ \bibinfo {author} {\bibfnamefont {J.}~\bibnamefont
  {Silk}},\ }\href {\doibase 10.1038/s41550-019-0906-9} {\bibfield  {journal}
  {\bibinfo  {journal} {Nature Astron.}\ }\textbf {\bibinfo {volume} {4}},\
  \bibinfo {pages} {196} (\bibinfo {year} {2019})},\ \Eprint
  {http://arxiv.org/abs/1911.02087} {arXiv:1911.02087 [astro-ph.CO]}
  \BibitemShut {NoStop}%
\bibitem [{\citenamefont {Efstathiou}\ and\ \citenamefont
  {Gratton}(2020)}]{Efstathiou:2020wem}%
  \BibitemOpen
  \bibfield  {author} {\bibinfo {author} {\bibfnamefont {G.}~\bibnamefont
  {Efstathiou}}\ and\ \bibinfo {author} {\bibfnamefont {S.}~\bibnamefont
  {Gratton}},\ }\href {\doibase 10.1093/mnrasl/slaa093} {\bibfield  {journal}
  {\bibinfo  {journal} {Mon. Not. Roy. Astron. Soc.}\ }\textbf {\bibinfo
  {volume} {496}},\ \bibinfo {pages} {L91} (\bibinfo {year} {2020})},\ \Eprint
  {http://arxiv.org/abs/2002.06892} {arXiv:2002.06892 [astro-ph.CO]}
  \BibitemShut {NoStop}%
\bibitem [{\citenamefont {Metropolis}\ \emph {et~al.}(1953)\citenamefont
  {Metropolis}, \citenamefont {Rosenbluth}, \citenamefont {Rosenbluth},
  \citenamefont {Teller},\ and\ \citenamefont {Teller}}]{Metropolis:1953}%
  \BibitemOpen
  \bibfield  {author} {\bibinfo {author} {\bibfnamefont {N.}~\bibnamefont
  {Metropolis}}, \bibinfo {author} {\bibfnamefont {A.}~\bibnamefont
  {Rosenbluth}}, \bibinfo {author} {\bibfnamefont {M.}~\bibnamefont
  {Rosenbluth}}, \bibinfo {author} {\bibfnamefont {A.}~\bibnamefont {Teller}},
  \ and\ \bibinfo {author} {\bibfnamefont {E.}~\bibnamefont {Teller}},\ }\href
  {\doibase 10.1063/1.1699114} {\bibfield  {journal} {\bibinfo  {journal}
  {Journal of Chemical Physics}\ }\textbf {\bibinfo {volume} {21}},\ \bibinfo
  {pages} {1087} (\bibinfo {year} {1953})}\BibitemShut {NoStop}%
\bibitem [{\citenamefont {Hastings}(1970)}]{Hastings:1970}%
  \BibitemOpen
  \bibfield  {author} {\bibinfo {author} {\bibfnamefont {W.}~\bibnamefont
  {Hastings}},\ }\href {\doibase 10.1093/biomet/57.1.97} {\bibfield  {journal}
  {\bibinfo  {journal} {Biometrika}\ }\textbf {\bibinfo {volume} {57}},\
  \bibinfo {pages} {97} (\bibinfo {year} {1970})}\BibitemShut {NoStop}%
\bibitem [{\citenamefont {Lewis}(2019)}]{Lewis:2019xzd}%
  \BibitemOpen
  \bibfield  {author} {\bibinfo {author} {\bibfnamefont {A.}~\bibnamefont
  {Lewis}},\ }\href {https://getdist.readthedocs.io} {\  (\bibinfo {year}
  {2019})},\ \Eprint {http://arxiv.org/abs/1910.13970} {arXiv:1910.13970
  [astro-ph.IM]} \BibitemShut {NoStop}%
\bibitem [{\citenamefont {Renzi}\ and\ \citenamefont
  {Silvestri}(2020)}]{Renzi:2020fnx}%
  \BibitemOpen
  \bibfield  {author} {\bibinfo {author} {\bibfnamefont {F.}~\bibnamefont
  {Renzi}}\ and\ \bibinfo {author} {\bibfnamefont {A.}~\bibnamefont
  {Silvestri}},\ }\href@noop {} {\  (\bibinfo {year} {2020})},\ \Eprint
  {http://arxiv.org/abs/2011.10559} {arXiv:2011.10559 [astro-ph.CO]}
  \BibitemShut {NoStop}%
\bibitem [{\citenamefont {Macpherson}\ and\ \citenamefont
  {Heinesen}(2021)}]{Macpherson:2021gbh}%
  \BibitemOpen
  \bibfield  {author} {\bibinfo {author} {\bibfnamefont {H.~J.}\ \bibnamefont
  {Macpherson}}\ and\ \bibinfo {author} {\bibfnamefont {A.}~\bibnamefont
  {Heinesen}},\ }\href {\doibase 10.1103/PhysRevD.104.109901} {\  (\bibinfo
  {year} {2021}),\ 10.1103/PhysRevD.104.109901},\ \bibinfo {note} {[Erratum:
  Phys.Rev.D 104, 109901 (2021)]},\ \Eprint {http://arxiv.org/abs/2103.11918}
  {arXiv:2103.11918 [astro-ph.CO]} \BibitemShut {NoStop}%
\bibitem [{\citenamefont {Krishnan}\ \emph {et~al.}(2021)\citenamefont
  {Krishnan}, \citenamefont {Mohayaee}, \citenamefont {Colg\'ain},
  \citenamefont {Sheikh-Jabbari},\ and\ \citenamefont
  {Yin}}]{Krishnan:2021dyb}%
  \BibitemOpen
  \bibfield  {author} {\bibinfo {author} {\bibfnamefont {C.}~\bibnamefont
  {Krishnan}}, \bibinfo {author} {\bibfnamefont {R.}~\bibnamefont {Mohayaee}},
  \bibinfo {author} {\bibfnamefont {E.~O.}\ \bibnamefont {Colg\'ain}}, \bibinfo
  {author} {\bibfnamefont {M.~M.}\ \bibnamefont {Sheikh-Jabbari}}, \ and\
  \bibinfo {author} {\bibfnamefont {L.}~\bibnamefont {Yin}},\ }\href {\doibase
  10.1088/1361-6382/ac1a81} {\bibfield  {journal} {\bibinfo  {journal} {Class.
  Quant. Grav.}\ }\textbf {\bibinfo {volume} {38}},\ \bibinfo {pages} {184001}
  (\bibinfo {year} {2021})},\ \Eprint {http://arxiv.org/abs/2105.09790}
  {arXiv:2105.09790 [astro-ph.CO]} \BibitemShut {NoStop}%
\bibitem [{\citenamefont {Rasmussen}\ and\ \citenamefont
  {Williams}(2006)}]{RasmussenWilliams}%
  \BibitemOpen
  \bibfield  {author} {\bibinfo {author} {\bibfnamefont {C.}~\bibnamefont
  {Rasmussen}}\ and\ \bibinfo {author} {\bibfnamefont {C.}~\bibnamefont
  {Williams}},\ }\href@noop {} {\emph {\bibinfo {title} {Gaussian Processes for
  Machine Learning}}}\ (\bibinfo  {publisher} {MIT Press},\ \bibinfo {address}
  {Massachusetts},\ \bibinfo {year} {2006})\BibitemShut {NoStop}%
\bibitem [{\citenamefont {Busti}\ \emph {et~al.}(2014)\citenamefont {Busti},
  \citenamefont {Clarkson},\ and\ \citenamefont {Seikel}}]{Busti:2014dua}%
  \BibitemOpen
  \bibfield  {author} {\bibinfo {author} {\bibfnamefont {V.~C.}\ \bibnamefont
  {Busti}}, \bibinfo {author} {\bibfnamefont {C.}~\bibnamefont {Clarkson}}, \
  and\ \bibinfo {author} {\bibfnamefont {M.}~\bibnamefont {Seikel}},\ }\href
  {\doibase 10.1093/mnrasl/slu035} {\bibfield  {journal} {\bibinfo  {journal}
  {Mon. Not. Roy. Astron. Soc.}\ }\textbf {\bibinfo {volume} {441}},\ \bibinfo
  {pages} {11} (\bibinfo {year} {2014})},\ \Eprint
  {http://arxiv.org/abs/1402.5429} {arXiv:1402.5429 [astro-ph.CO]} \BibitemShut
  {NoStop}%
\bibitem [{\citenamefont {Verde}\ \emph {et~al.}(2014)\citenamefont {Verde},
  \citenamefont {Protopapas},\ and\ \citenamefont {Jim\'enez}}]{Verde:2014qea}%
  \BibitemOpen
  \bibfield  {author} {\bibinfo {author} {\bibfnamefont {L.}~\bibnamefont
  {Verde}}, \bibinfo {author} {\bibfnamefont {P.}~\bibnamefont {Protopapas}}, \
  and\ \bibinfo {author} {\bibfnamefont {R.}~\bibnamefont {Jim\'enez}},\ }\href
  {\doibase 10.1016/j.dark.2014.09.003} {\bibfield  {journal} {\bibinfo
  {journal} {Phys. Dark Univ.}\ }\textbf {\bibinfo {volume} {5-6}},\ \bibinfo
  {pages} {307} (\bibinfo {year} {2014})},\ \Eprint
  {http://arxiv.org/abs/1403.2181} {arXiv:1403.2181 [astro-ph.CO]} \BibitemShut
  {NoStop}%
\bibitem [{\citenamefont {Yu}\ \emph {et~al.}(2018)\citenamefont {Yu},
  \citenamefont {Ratra},\ and\ \citenamefont {Wang}}]{Yu:2017iju}%
  \BibitemOpen
  \bibfield  {author} {\bibinfo {author} {\bibfnamefont {H.}~\bibnamefont
  {Yu}}, \bibinfo {author} {\bibfnamefont {B.}~\bibnamefont {Ratra}}, \ and\
  \bibinfo {author} {\bibfnamefont {F.-Y.}\ \bibnamefont {Wang}},\ }\href
  {\doibase 10.3847/1538-4357/aab0a2} {\bibfield  {journal} {\bibinfo
  {journal} {Astrophys. J.}\ }\textbf {\bibinfo {volume} {856}},\ \bibinfo
  {pages} {3} (\bibinfo {year} {2018})},\ \Eprint
  {http://arxiv.org/abs/1711.03437} {arXiv:1711.03437 [astro-ph.CO]}
  \BibitemShut {NoStop}%
\bibitem [{\citenamefont {G\'omez-Valent}\ and\ \citenamefont
  {Amendola}(2018)}]{Gomez-Valent:2018hwc}%
  \BibitemOpen
  \bibfield  {author} {\bibinfo {author} {\bibfnamefont {A.}~\bibnamefont
  {G\'omez-Valent}}\ and\ \bibinfo {author} {\bibfnamefont {L.}~\bibnamefont
  {Amendola}},\ }\href {\doibase 10.1088/1475-7516/2018/04/051} {\bibfield
  {journal} {\bibinfo  {journal} {JCAP}\ }\textbf {\bibinfo {volume} {04}},\
  \bibinfo {pages} {051} (\bibinfo {year} {2018})},\ \Eprint
  {http://arxiv.org/abs/1802.01505} {arXiv:1802.01505 [astro-ph.CO]}
  \BibitemShut {NoStop}%
\end{thebibliography}%

\end{document}